\documentclass{article}

\usepackage{arxiv}

\usepackage[utf8]{inputenc} 
\usepackage[T1]{fontenc}    
\usepackage{hyperref}       
\usepackage{url}            
\usepackage{booktabs}       
\usepackage{amsfonts}       
\usepackage{nicefrac}       
\usepackage{microtype}      
\usepackage{graphicx}
\usepackage{natbib}
\usepackage{doi}

\usepackage{amsmath}
\usepackage{tabularx}
\usepackage{interval}

\title{Role of Diversity in Team Performance: the Case of Missing Expertise, an Agent Based Simulation}

\author{\href{https://orcid.org/0000-0001-6360-0714}{\includegraphics[scale=0.06]{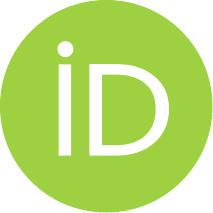}\hspace{1mm}Tam\'as Kiss}\thanks{Webpage: \url{https://tncs.wigner.hu/people/kiss/}} \\
	Department of Computational Sciences\\
	Wigner Research Centre for Physics\\
	Hungarian Research Network\\
	\texttt{kiss.t@wigner.hu}\\
}

\date{}

\renewcommand{\shorttitle}{Role of Diversity in Team Performance}

\hypersetup{
pdftitle={Role of Diversity in Team Performance: the Case of Missing Expertise, an Agent Based Simulation},
pdfsubject={cs.MA, physics.soc-ph},
pdfauthor={Tam\'as Kiss},
pdfkeywords={Functional Diversity, Agent-Based Simulations, Organizational Design, Management Teams, Communication},
}

\newcommand{\argmaxT}[1]{\underset{#1}{\mathrm{argmax}\ }}

\begin{document}
\maketitle

\begin{abstract}
Theory and empirical research on management teams' influence on firm performance have witnessed continuous development, and by now incorporate numerous details. Classic, experiment-based studies examining social systems collect vast amount of data, but often times investigate only the first one or two modes of the distribution of measured variables, and experience difficulty in analyzing the effect of context. For example, in functional diversity research, management teams are described by measures incorporating complex distributions of capabilities of individual managers and teams of managers. To investigate the effect of hidden distributions, and the effect of functional diversity composition on team communication and performance, we developed an agent-based model, and conducted a series of simulation experiments. Modeling results show that depending on the context, such as communication scheme among interacting agents, or their functional composition, intrapersonal functional diversity (IFD), and dominant function diversity (DFD) might enhance or reduce performance and communication among agents. Furthermore, simulation results also suggest that a third measure is required alongside IFD and DFD capturing the aggregate expertise of the team to comprehensively account for empirical findings.
\end{abstract}


\section{Introduction}\label{sec:intro}

Corporate management teams influence firm performance through a number of ways, including for example strategic planning and decision-making \citep{hambrick_upper_1984}, and via social psychological processes \citep{jehn_why_1999,van_knippenberg_work_2004}. The ability of a management team to bring good decisions, in turn, largely depends on the composition of the team, and interaction of its members with each other, and the company’s environment. All of these constructs -- firm performance, team composition, and member interactions – have been defined and assayed empirically in a number of studies \citep{aboramadan_top_2021}.  Management theory highlights the importance of diversity in the definition of team composition \citep{williams_demography_1998,cannella_top_2008}, and communication in member interactions \citep{smith_top_1994,de_carolis_why_2009}. To describe team composition, an experimentally well tractable measure, functional diversity in teams, has been proposed, and formulated for example as ``the occupational backgrounds and functional areas of expertise of the team members'' \citep{jackson_consequences_1996}.

\citet{bunderson_comparing_2002} showed that functional diversity in teams can be conceptualized and measured in a number of different ways. Conscious definition and precise understanding of functional diversity measures are of great importance since empirical research on the role of functional diversity in influencing team effectiveness has shown a complex picture, and often resulted in contradictory findings \citep{homberg_top_2013}.  In particular, \citet{bunderson_comparing_2002} compared the effect of intrapersonal functional diversity (IFD) and dominant function diversity (DFD) on information sharing within a team, and on near-term performance of a business unit. They showed that IFD is positively associated with information sharing, and information sharing mediates the positive relationship between IFD and performance.  Moreover, they also showed that DFD is negatively associated with information sharing, and information sharing partially mediates the negative relationship between DFD and performance.

Interestingly enough, a recent study by \citet{zhou_functional_2023}, using the same definition and empirical assessment of IFD and DFD found that both of these measures have positive effect on performance of a firm. Comparison of experimental findings shows, however, that it is very difficult to identify the effect of a certain independent variable, such as DFD, because the influence of moderator variables might be overwhelming. Furthermore, the domain in which experimental studies can measure certain variables might largely constrain observable effects. For example, while mean values of IFD and DFD were 0.28 and 0.66, respectively in the study by \citeauthor{bunderson_comparing_2002}, \citeauthor{zhou_functional_2023} reported the same values to be 0.58, and 0.59, respectively (IFD and DFD values range between 0 and 1). Indeed, \citeauthor{bunderson_comparing_2002} noted that ``the seemingly low intrapersonal functional diversity scores for these business unit management teams [\dots] suggests that, for the most part, managers in this sample had fairly narrow ranges of functional experience''.

A number of researchers have previously developed models with the aim of reflecting some aspect of human behavior or teamwork \citep{hong_groups_2004, chang_differential_2012, dehkordi_impacts_2012, vermillion_using_2015, fernandes_modelling_2017}. For example, applying a formal approach, \citet{hong_groups_2004} developed a mathematical theorem and by showing a trade-off between diversity and ability –- in the limit of large number of problem solvers –- they proved that a random group of diverse problem solvers outperform the group of best problem solvers. \citet{chang_differential_2012} used an agent-based model (ABM; \citet{taylor_agent-based_2014}) to show that two specific types of knowledge diversity dimensions, intrapersonal knowledge diversity, and shared knowledge diversity have different effects on team innovation: While intrapersonal knowledge diversity is positively related to team innovation, shared knowledge diversity is not related to it. \citet{dehkordi_impacts_2012} modeled the effect of stress and motivation on team performance, and studied the impact of project overload. They showed that work overload stifles innovation, while motivation on challenging goals might alleviate the negative aspects of project overload. \citet{vermillion_using_2015} used agent-based modeling approaches to investigate the delegation of authority and use of incentives in design teams, and \citet{fernandes_modelling_2017} developed an ABM to support design teams of industrial organizations in understanding complex cause–effect relationships in early design projects.

While a large body of literature exists to address specific questions or problems, a number of general-purpose simulators also exist to conduct agent-based modeling. For example, the Virtual Design Team (or VDT) model by \citet{jin_virtual_1996} explicitly models actors, activities and organizations, and its goal is to simulate and analyze how activity interdependencies raise coordination needs. Varying model parameters describing communication tools sheds light on their effect on team performance. Simulation tools, like TCM \citep{rojas-villafane_agent-based_2010} and NetWatch \citep{tsvetovat_modeling_2004} similarly represent individuals possessing human characteristics like motivation, personal traits, memory or learning ability, and can be used to predict team performance. The main purpose of systems like, for example, GRATE* \citep{jennings_controlling_1995}, STEAM \citep{tambe_towards_1997}, CAST \citep{yen_cast_2001}, and a recent model by \citet{perisic_agent-based_2016} is to simulate teamwork behaviors in order to improve team effectiveness. For reviews on using ABMs of human systems, see for example the works by \citet{bonabeau_agent-based_2002} and \citet{fan_modeling_2004}.

In this study, an ABM was developed, and simulation experiments were conducted to study effects of empirically hard-to-control moderator variables and functions. We highlight the importance of viewing management processes as complex, emergent phenomena shaped by diversity \citep{nkomo_diversity_2019}, and note that some statistical methods used to analyze empirically collected data employ a number of projections, and hence might miss the fine details hidden in this data. For example, IFD \citep{walsh_selectivity_1988} and DFD \citep{hirschman_paternity_1964} span a two-dimensional space in which dependent variables, like the amount of communication or team performance potentially form non-linear, exotic surfaces. While empirical studies typically account for the interaction effect in regression-type analyses, local deviations from the large-scale behavior might still be missed. Using ABMs and detailed data visualization could shed light to new, exciting phenomena.

The specific goal of this work is to elucidate on the effect of DFD on team performance which, in some circumstances positively influence team performance, while in others has a negative effect on it. These efforts led to the suggestion that measurement of IFD and DFD alone does not fully capture diversity properties of even simple problem-solving systems, hence we propose to introduce a new diversity measure describing the aggregate knowledge base available to a management team as a whole.

\section{System Modelling}\label{sec:model}

\subsection{Model components}\label{ssec:components}
An ABM was developed to simulate task processing and communication in a group of interacting agents representing managers of management teams. This computer model consists of tasks, agents and their interactions, and can be parameterized to simulate a number of scenarios. 

\begin{figure}[ht]
\centering
\includegraphics[width=\textwidth]{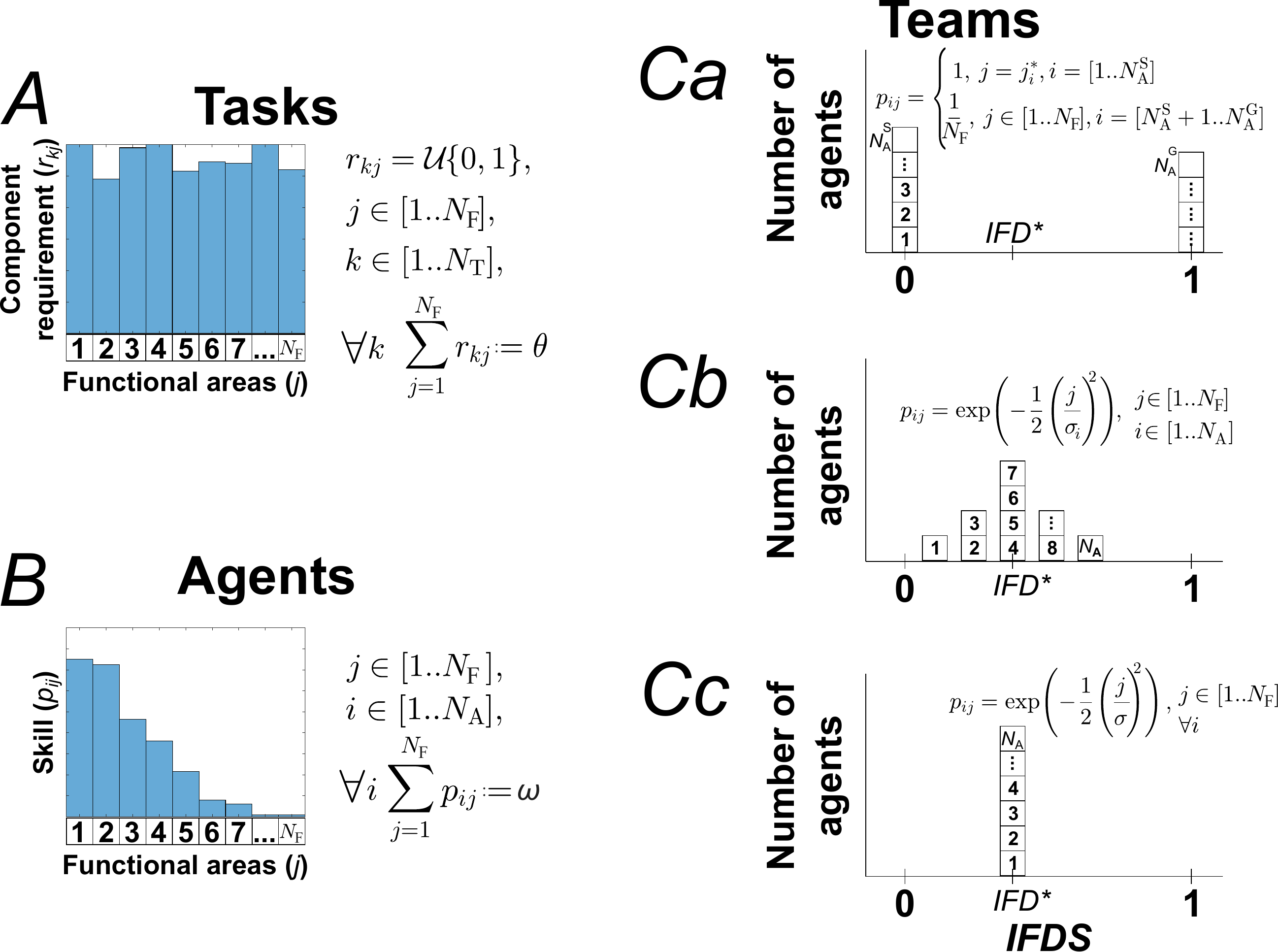}
\caption{Representation of tasks (\textit{A}) and agents (\textit{B}) in the model. $r_{kj}$ denotes the amount of work needed to complete component $j$ of task $k$ and is drawn from a uniform random distribution. $p_{ij}$ denotes strength of skill $j$ of agent $i$. $N_\mathrm{F}$ is the number of functional components, $N_\mathrm{T}$ and $N_\mathrm{A}$ are the number of tasks and agents, respectively. \textit{C}: Approaches to generate groups of agents using different individual functional diversity score (IFDS; Equation\ \ref{eq:IFDS}) distributions. IFD values (Equation\ \ref{eq:IFD}) of the groups shown in the above examples are all identical ($\mathit{IFD}^*$) even though teams are composed of agents with different IFDS distributions. \textit{Ca}: The pure specialist–-absolute generalist system with $N_\mathrm{A}^\mathrm{S}$ specialists having $\mathit{IFDS} = 0$, and $N_\mathrm{A}^\mathrm{G}$ generalists having $\mathit{IFDS} = 1$. \textit{Cb}: a system of agents with IFDS distributed according to a Gaussian distribution around $\mathit{IFD}^*$. In this case agents are described by $p_{ij}$ skill distributions following a discrete half-normal distribution of standard deviation $\sigma_i$. \textit{Cc}: A system with agents having identical IFDS values, all equaling $\mathit{IFD}^*$.}\label{fig:TaskAgTeam}
\end{figure}

Each task $k$ (Figure \ref{fig:TaskAgTeam} \textit{A}) is said to be composed of $N_\mathrm{Functions}$ components representing different functions, and each component $j$ requires some amount of ``work'', denoted by $r_{kj}$, to be completed. In this representation, each task encompasses a complete production workflow that requires multiple skills or functions, like marketing, manufacturing, sales, distribution, etc. In particular, the $r_{kj}$ values are drawn from a uniform random distribution for each task, and their sum is normalized to a fixed value $\theta$. This way, tasks are defined with different amount of work required to complete each of their component, but having the same total work requirement. A task is said to be completed when all of its components are completed. 

Agents are similarly generated: each agent is defined by $N_\mathrm{Functions}$ ``skills'', and these skills are normalized to $\omega$, representing agents who work the same amount, but have a different set of capabilities (Figure \ref{fig:TaskAgTeam} \textit{B}). For each agent $i$, the $p_{ij}$ strength value for skill $j$ is randomly generated or specifically assigned, depending on parameter $\nu$ (Table \ref{tab:params}), representing the mode of agent generation. Specifically, for simulations of a team of pure specialists and absolute generalists (Figure \ref{fig:TaskAgTeam} \textit{Ca}), a randomly selected skill ($j^*_i$) was set to have strength value $p_{ij^*_i} = \omega$, and all other skills had $0$ strength for specialists, or all skill strength values were set to $\frac{\omega}{N_\mathrm{Functions}}$ for generalists. For simulations of other team types, $p_{ij}$ skill strength values were drawn from a half-normal distribution with standard deviation set for each agent independently (Figure \ref{fig:TaskAgTeam} \textit{Cb}), or identically (Figure \ref{fig:TaskAgTeam} \textit{Cc}).

Functional diversity of an agent's skill set is described by the individual functional diversity score or IFDS \citep{walsh_selectivity_1988}. It is a normalized measure of how diverse an agent’s functional expertise is, and 0 represents a specialist with knowledge in only one functional area, while 1 represents a generalist with the same amount of knowledge across all functional areas. Specifically, for each agent $i$ $\mathit{IFDS}_i$ is defined as
\begin{equation}
    \mathit{IFDS}_i = \frac{1-\sum\limits_{j=1}^{N_\mathrm{Functions}} p_{ij}^2}{\rho}
    \label{eq:IFDS}
\end{equation}
where $j$ indexes functional areas, and $\rho=1-\frac{1}{N_\mathrm{Functions}}$ is a normalizing factor to scale IFDS into the $\interval{0}{1}$ interval.

Teams are composed of $N_\mathrm{Agents}$ agents, and are characterized by the two functional diversity measures IFD, the average of individual IFDS values, and DFD as
\begin{align}
    \mathit{IFD} &= \frac{1}{N_\mathrm{Agents}}\sum\limits_{i=1}^{N_\mathrm{Agents}} \mathit{IFDS}_i\label{eq:IFD}\\
    \mathit{DFD} &= \frac{1-\sum\limits_{j=1}^{N_\mathrm{Functions}} \left(\frac{\argmaxT{j} p_{ij}}{N_\mathrm{Agents}}\right)^2}{\rho}
    \label{eq:DFD}
\end{align}
where $\argmaxT{j} p_{ij}$ specifies the strongest skill of agent $i$. It is important to note, however, that even though skill strength value distributions might largely differ across agents of different teams, their IFD, and DFD measures could still be identical (Figure \ref{fig:TaskAgTeam} \textit{C}). For example, as an extreme example, agents in Group \textit{Ca} are either specialists or generalists with $\mathit{IFDS}_i$ equal to 0 or 1, respectively, yet, their IFD value might be the same as the IFD of the following groups, i.e.\ $\mathit{IFD^*}$. Indeed, the average of $\mathit{IFDS}_i$ of Group \textit{Cb} might also yield the same $\mathit{IFD^*}$, even though agents are described by different $\mathit{IFDS}_i$ values. Furthermore, Group \textit{Cc} (not used in presented simulations) represents a case when all agents have identically parameterized $p_{ij}$ distribution independent of the agent's index $i$, resulting in the same $\mathit{IFDS}$ values, yielding an $\mathit{IFD^*}$ equal to this $\mathit{IFDS}$. Hence, using only IFD and DFD to describe a managerial group hides the information about properties of individual actors, which might influence performance of the team as a whole.

\begin{figure}[ht]
    \centering
    \includegraphics[width=\textwidth]{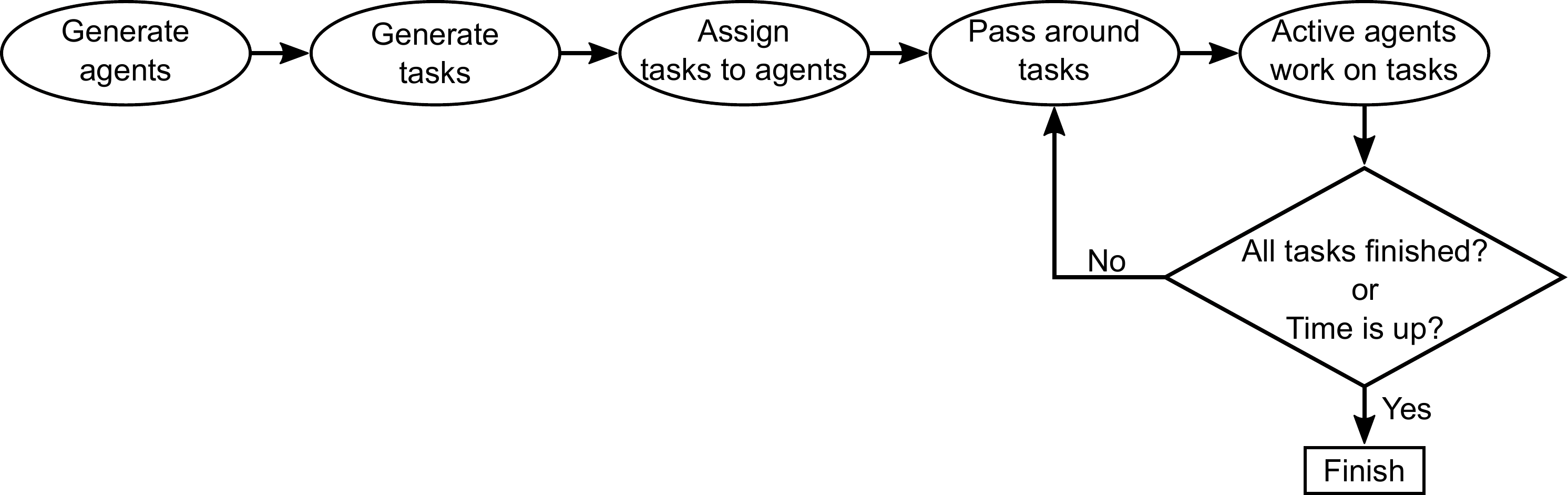}
    \caption{Model Dynamics. The steps shown in the figure are iterated $r$ times for given IFD and DFD to generate $r$ instances of the system allowing the calculation of average communication and performance values.}
    \label{fig:moddyn}
\end{figure}

\subsection{System dynamics, and communication among agents}
Dynamics of the system (Figure \ref{fig:moddyn}) is composed of the following steps:  First, agents and tasks are generated by creating $p_{ij}$ agent skill vectors and $r_{kj}$ task component vectors, respectively. Second, tasks are assigned to agents randomly. In any time step, an agent might have zero or one task assigned to them. If there are fewer tasks than agents, some agents are idle, if there are more tasks than agents, some tasks are unassigned and do not progress towards completion. Third, agents might pass their task to another agent as detailed below. Fourth, agents work on the task assigned to them. Here, work is simulated by subtracting the skill vector of an agent from the component vector of the task assigned to the agent. For example, the work agent $i$ performs on the $j^\mathrm{th}$ component of task $k$ is expressed as $w_{ijk} = r_{kj}-p_{ij}$. An agent performs operations on all non-completed task components simultaneously. Once an $r_{kj}$ value reaches $0$ the component is considered completed and does not decrease any further. If all components of a task are completed, the task itself is completed, removed from the simulation and the assigned agent becomes idle. If all tasks are completed, the simulation finishes. The simulation also finishes if a pre-defined number of simulation steps ($M$) are taken, since there are situations when agents are not able to solve all tasks due to an inadequate skill set or lack of communication. If there are unsolved tasks and available time to work on the tasks, the simulation continues with passing tasks again.\label{ssec:dyna}

Communication in this model is simulated by passing a task. Task passing represents the process of team members shifting the focus of the production workflow from one function to another, when a function is completed, or the processing agent cannot make progress anymore. The passing of a task is constrained by similarity of the agents. Specifically, the distance ($d_{mn}$) between agents $m$ and $n$ is defined as the Euclidean distance between their skill vectors as
\begin{equation}
    d_{mn} = \sqrt{\sum\limits_{j=1}^{N_\mathrm{Functions}} (p_{mj} - p_{nj})^2}.
    \label{eq:dist}
\end{equation}
Only agents closer to each other than a threshold ($d_{mn} < \tau$) are allowed to pass and receive tasks from each other. Such agents will be termed as collaborators.\label{ssec:comm}

\begin{figure}[ht]
    \centering
    \includegraphics[width=\textwidth]{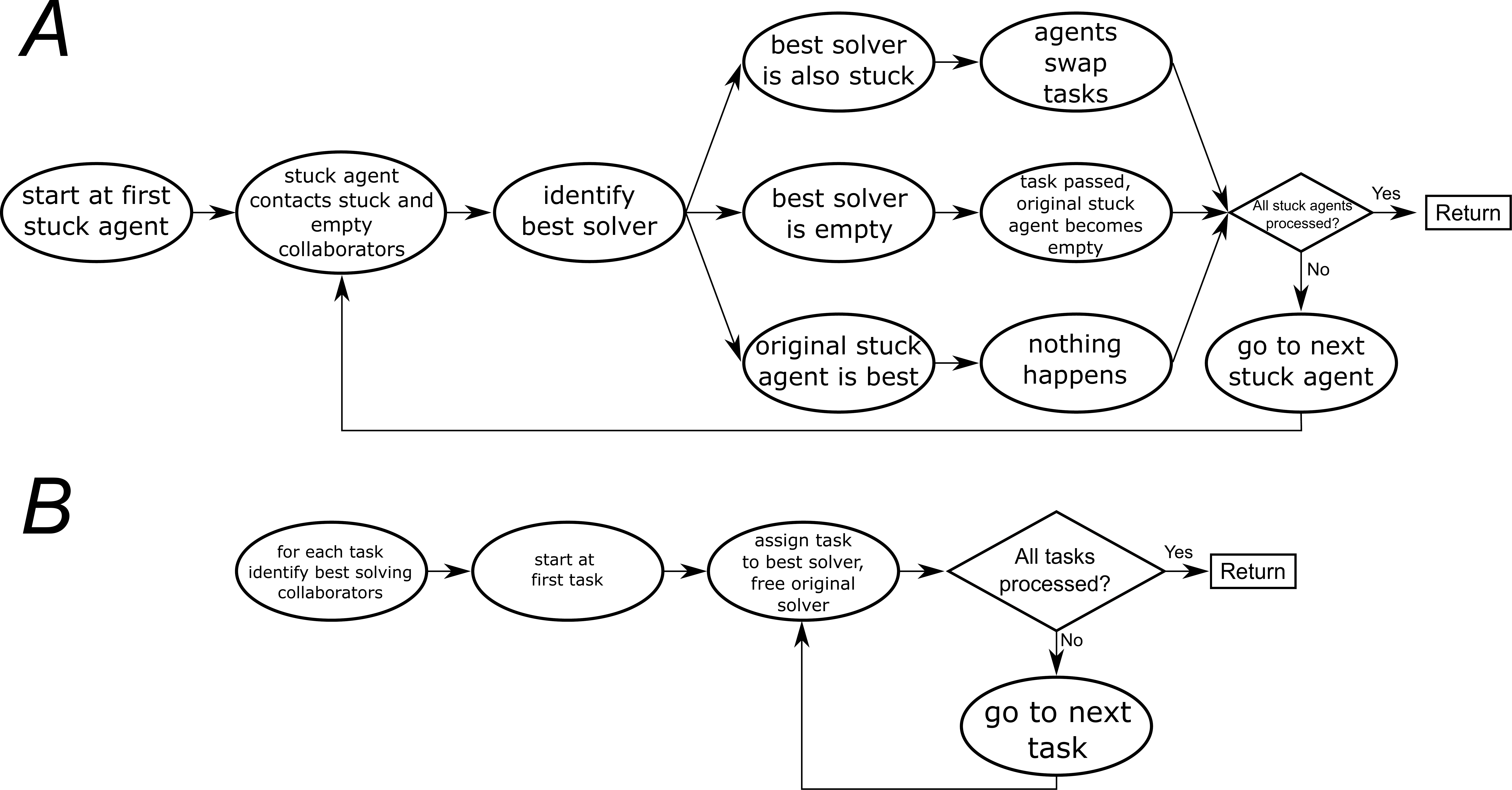}
    \caption{Communication Schemes. Communication in this ABM is represented by task passing. \textit{A}: In the first scheme, agents only look for a better solver among their collaborator if they cannot proceed with solving their task (agent is ``stuck''). \textit{B}: In the second scheme, agents repeatedly seek to identify their collaborator who solves their task in the shortest time.}
    \label{fig:comm}
\end{figure}

To evaluate the effect of different communication schemes, two task passing rules were implemented. In the first version (Figure \ref{fig:comm} \textit{A}), when an agent cannot proceed with solving their task (get ``stuck''), evaluate how well any of their collaborators could proceed with their task, and pass the task to the one that is best in solving their task. In the second version (Figure \ref{fig:comm} \textit{B}), agents with a task decide once every simulation step whether to pass or keep their task, such that in every step they evaluate the amount of possible work they or any of their collaborators can perform in the next simulation step, and pass their task to the best performing collaborator, or keep it if they are the best solver for the given task.

\subsection{Model evaluation}\label{ssec:eval}
To help better understand the interaction between IFD and DFD, and to study behavior of teams composed of pure specialists and absolute generalists, communication and team performance will be shown in three-dimensional plots with IFD and DFD represented on the horizontal axes, and performance, communication, or other dependent variables color-coded on the vertical axis. Performance will be quantified by calculating the ratio of completed task components to the total number of task components initially introduced in the system, resulting in a value between $0$ and $1$. The amount of communication among agents will be represented by communication density, the number of task passes in the simulation normalized by the number of steps the simulation took.

\subsection{Summary of model parameters}

\begin{table}[h]
	\centering
	\begin{tabularx}{\textwidth}{lXl}
	\toprule
    Parameter & Description & Default value \\
	\midrule
    $N_\mathrm{Functions}$ & Number of functional domains accounted for in the simulations & 9\\
    $N_\mathrm{Agents}$ & Number of agents modeled & 10\\
    $N_\mathrm{Tasks}$ & Number of tasks modeled & 7\\
    $\omega$ & Agent normalization value, used to set total skill strength of an agent & 10\\
    $\Theta$ & Task normalization value. This parameter sets the total amount of effort required to complete a task & 10\\
    $\tau$ & Similarity threshold. When the skill strength vector of two agents are closer, in a Euclidean sense, then $\tau$, agents are allowed to exchange tasks & 80\%\\
    $\mu$ & Switch setting whether strength values of an agent are mixed across skills or skill strengths are serially generated & True\\
    $\nu$ & Specifies the method used to generate IFDS distribution of agents. Possible options are: same IFDS for all agents; IFDS distribution; only generalists and specialists (see Figure \ref{fig:TaskAgTeam} \textit{C} for details) & IFDS distribution\\
    $\delta$ & Width of IFDS distribution. Runs from 1 to $\frac{N_\mathrm{Agents}}{2}$ & $\frac{N_\mathrm{Agents}}{2}$\\
    $\pi$ & Specifies task passing scheme. Options are: always pass, pass if stuck & pass if stuck\\
    $r$ & Number of simulation instances for a given IFD -- DFD value & 10\\
    $M$ & Maximal number of time steps allowed & 250\\
    \bottomrule			
	\end{tabularx}
	\caption{List of parameters used in the model.}
	\label{tab:params}	
\end{table}

\subsection{Model implementation and dissemination}\label{ssec:implem}
For numerical simulations, IFD -- DFD value pairs were systematically scanned by generating a group of agents with given IFD and DFD, and running the computer model $r$ times to accumulate results for averaging. Simulation code was written in the Matlab\textcopyright\ language, and run on a personal computer. The full model is available on COMSeS Net and can be downloaded at: \url{https://www.comses.net/codebases/b5db6af8-ba44-4725-9bb3-09a6e6b02475/releases/1.0.0/}

\section{Results}\label{sec:res}

\begin{figure}[ht]
    \centering
    \includegraphics[width=\textwidth]{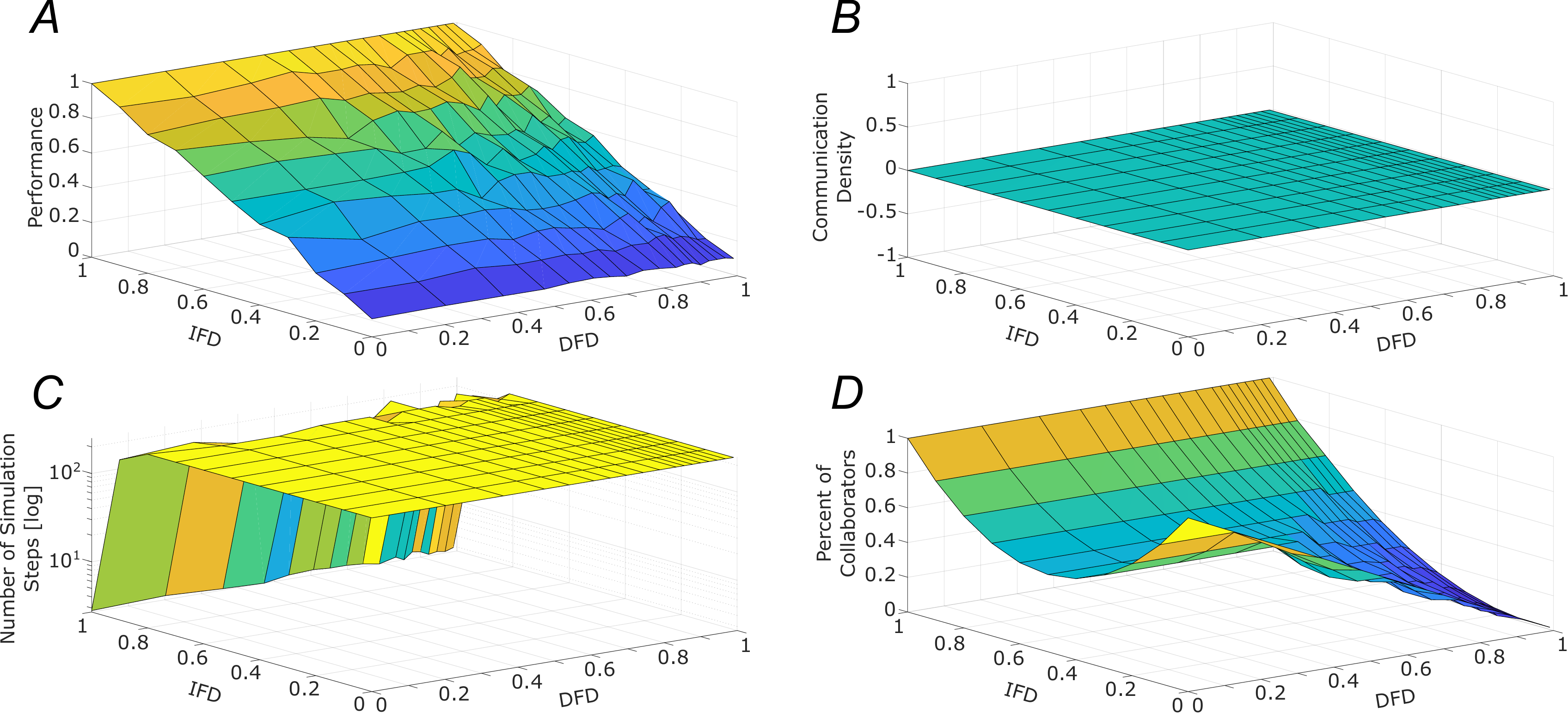}
    \caption{System behavior of teams of independently working specialists and generalists. \textit{A}: Performance of teams as a function of IFD and DFD. \textit{B}: Communication density of task passing among agents. \textit{C}: The number of simulation steps required to solve all tasks (on a log scale). $M=250$ time steps are the maximum allowed, even if there are unsolved tasks. \textit{D}: The number of collaborators relative to the number of all agent pairs in the team. Note that all tasks are solved in this system only when generalists are present. Also note that system behavior is independent of DFD.}
    \label{fig:specgen}
\end{figure}

\subsection{Pure specialists and absolute generalists: Answering a rhetorical question}\label{ssec:SpecGen}
In their \citeyear{bunderson_comparing_2002} article, \citeauthor{bunderson_comparing_2002} outline potential avenues for future research of the effect of diversity on teams, and propose that different types of functional diversity measures might interact with one another. They draw attention to the findings of \citet{jehn_why_1999}, who demonstrated that specific forms of team diversity and performance can be influenced by other forms of team diversity. This possibility of interaction prompts \citeauthor{bunderson_comparing_2002} to speculate on a hypothetical scenario of a management team consisting solely of specialists of the same function: "For example, the theory and results presented in this article raise an interesting paradox—what about a team composed entirely of specialists from the same function? On one hand, such a team should be able to easily share information (given a common functional background) but, on the other hand, increased information sharing may not translate into better informed decisions because the information shared represents a single functional perspective" \citep{bunderson_comparing_2002}.

\begin{figure}[!h]
    \centering
    \includegraphics[width=\textwidth]{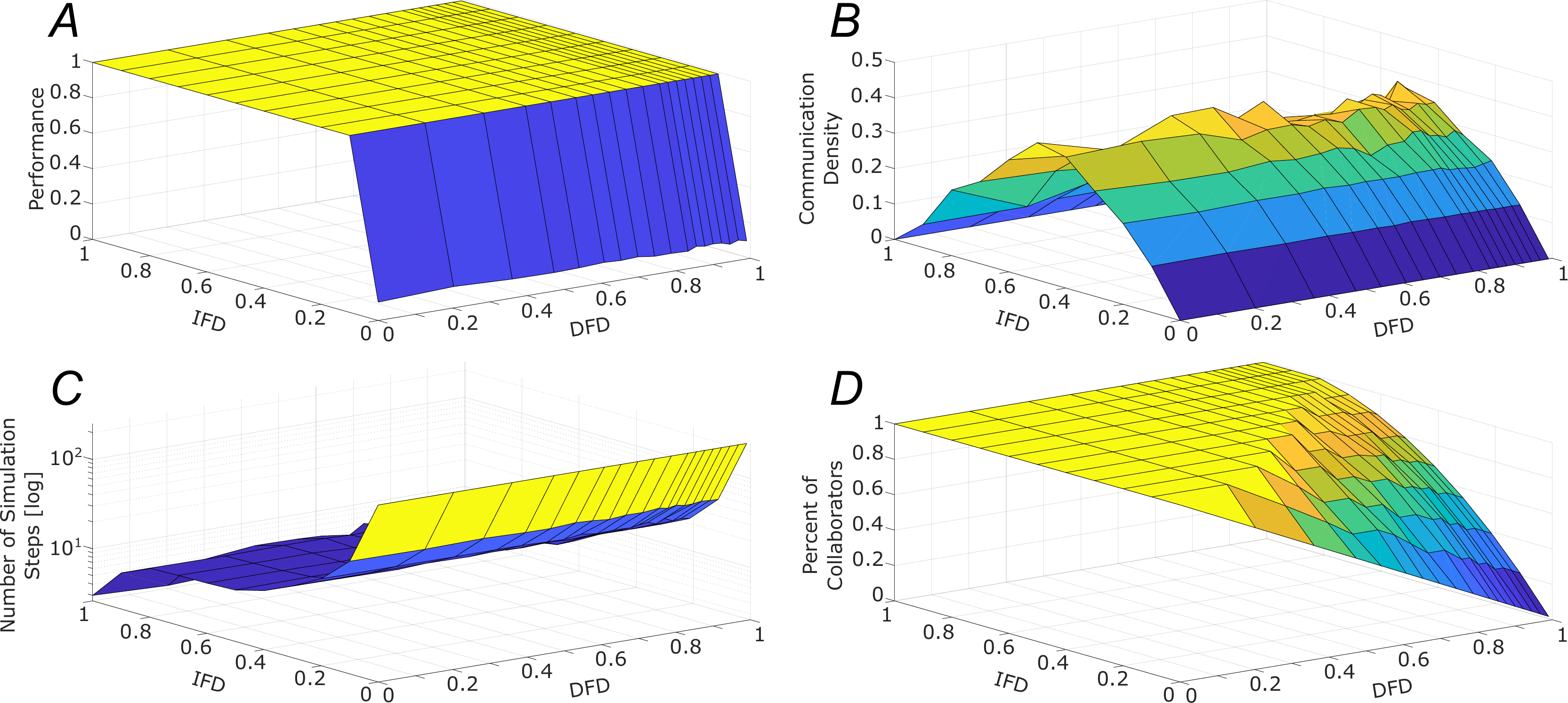}
    \caption{Simulation results of teams of communicating specialists and generalists. Figure panels are set up as in Figure \ref{fig:specgen}. Allowing communication among specialists and generalists results in superior performance, independent of the amount of communication and the value of DFD. For simulations, the first communication scheme was used, $\tau=80\%$ (see paragraph \ref{ssec:comm} for more details).}
    \label{fig:comspecgen}
\end{figure}

ABMs are particularly well-suited for examining theoretical scenarios like this, and this subsection is dedicated to exploring such a simple system. Following the above suggestion, the system consists of a mixture of pure specialists and absolute generalists (Figure \ref{fig:TaskAgTeam} \textit{Ca}). Pure specialists have only one skill (their IFDS is equal to 0), but they are very good in solving task components corresponding to their single skill. Absolute generalists, however, possess all possible skills (their IFDS is equal to 1) but they are not very good in any of those skills. In the model, values of a generalist’s skill strength add up to be equal to the value of a specialist’s single skill strength, representing problem solvers of about the same amount of total training or tenure length. IFD of a group is the average of all agents’ IFDS, and DFD is calculated using the strongest skill of agents, which is a random skill for generalists.

\begin{figure}[h]
    \centering
    \includegraphics[width=\textwidth]{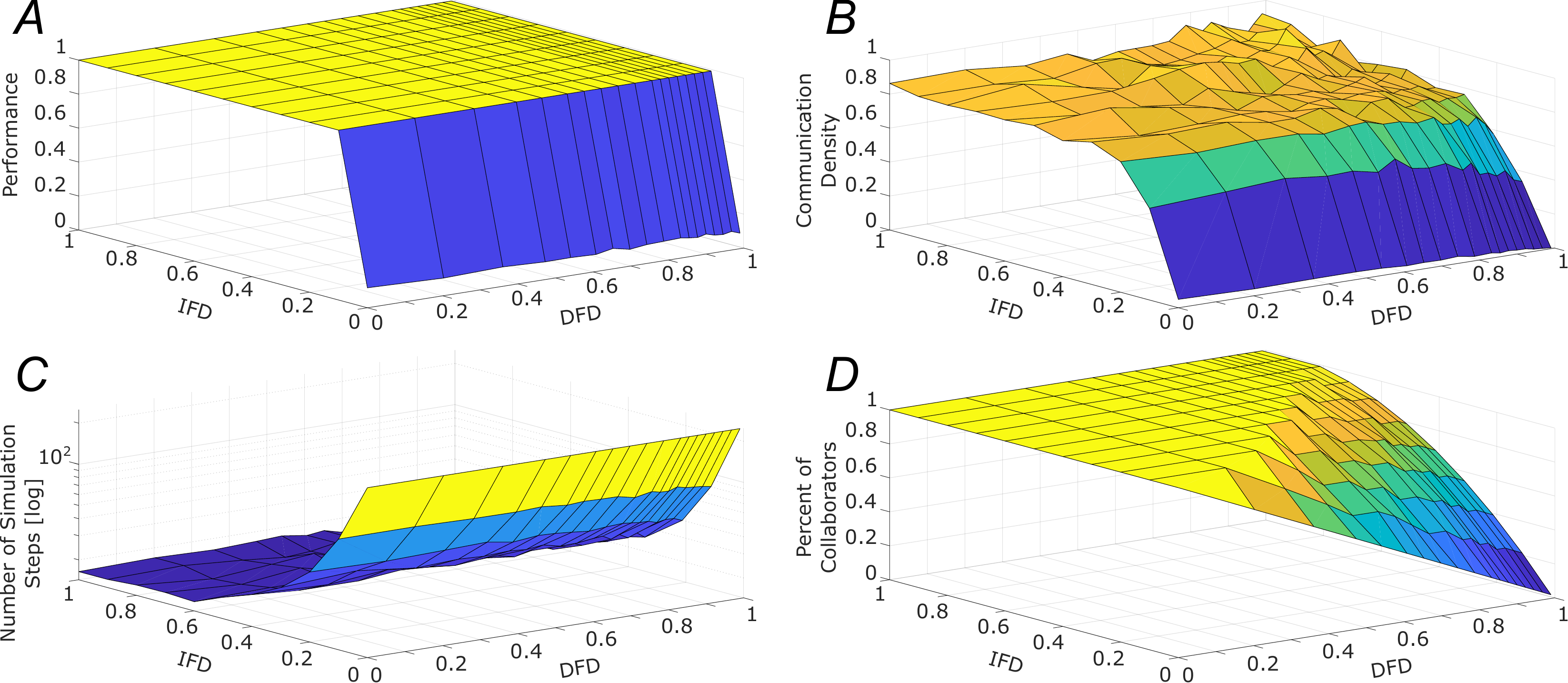}
    \caption{Simulation results of a system of generalists and specialists repeatedly seeking the best collaborator. Simulations using the second communication scheme, i.e.\ agents look for a more capable collaborator in every time step (c.f.\ Figure \ref{fig:comspecgen}). Figure panels are set up as in Figure \ref{fig:specgen}.}
    \label{fig:comsg2}
\end{figure}

This system can be in one of two possible states in terms of its performance, depending on the similarity threshold determining how similar agents must be to become collaborators and be capable of exchanging tasks. Figure \ref{fig:specgen} shows system behavior when the required similarity is too high to form hybrid collaborations, i.e.\ a specialist cannot communicate with a generalist, or specialists of different functions cannot communicate with each other, even though generalists can communicate with each other, similarly to specialists of the same function. When IFD and DFD are both 0, all team members are specialists of the same function, reflecting to the question of \citet{bunderson_comparing_2002}. Figure \ref{fig:specgen} \textit{A} shows that in this case the performance of the team is minimal. This is explained by the fact that agents are only able to solve the task component matching their single skill, all other components remain unsolved. Also, while all agents are identical, and as such are all collaborators (Figure \ref{fig:specgen} \textit{D}), they do not exchange tasks (Figure \ref{fig:specgen} \textit{B}), since communication does not bring new expertise into the team as predicted by \citeauthor{bunderson_comparing_2002}. Interestingly enough, performance does not increase with increasing DFD, i.e.\ introducing specialists of a different function, which clearly results from a lack of communication. In turn, this lack of communication is explained by the high similarity requirement disabling agents possessing different skills to communicate.

\begin{figure}[h]
    \centering
    \includegraphics[width=\textwidth]{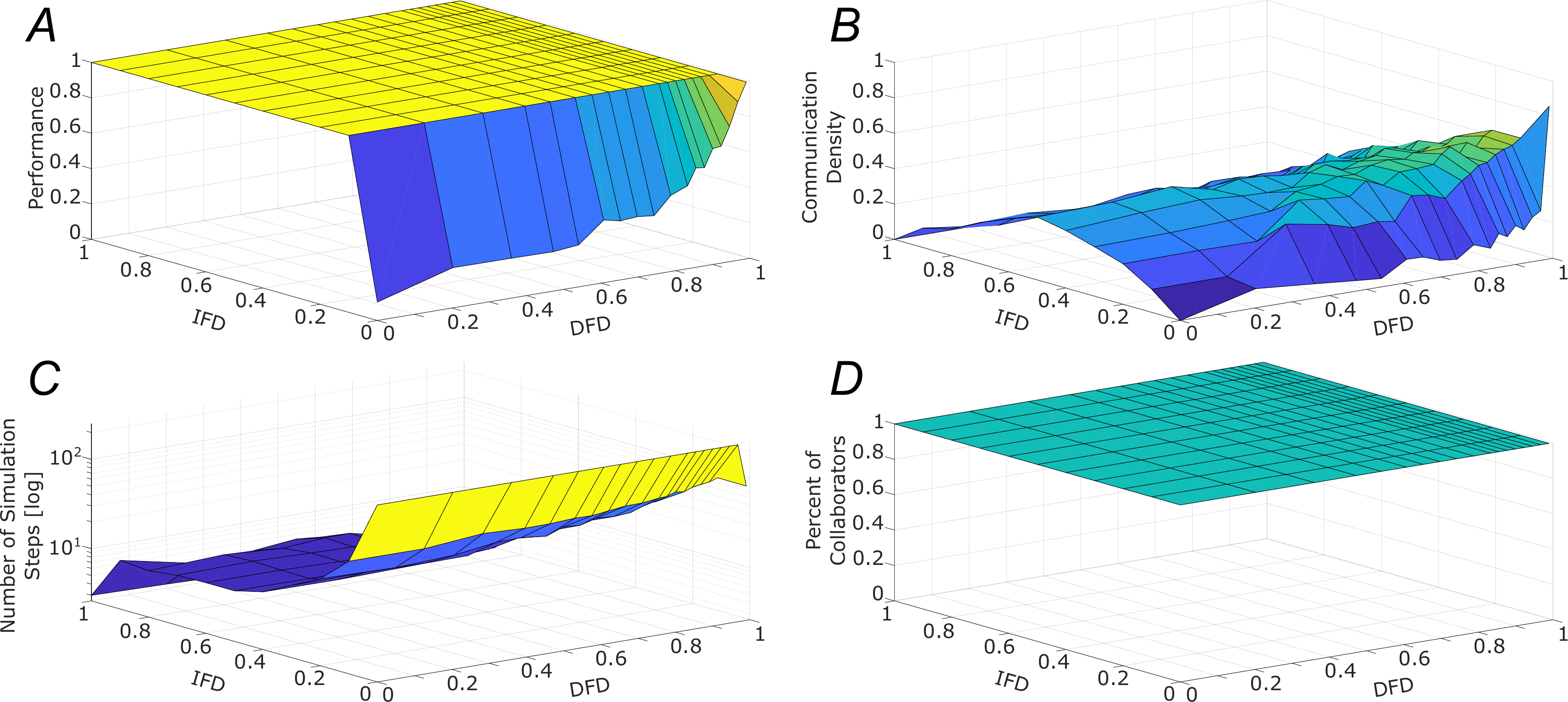}
    \caption{Simulation of generalists and specialists without communication constraints between agents. The first communication scheme was used in simulations, i.e.\ agents only look for collaborators to communicate with when they cannot proceed with solving their task. Figure panels are set up as in Figure \ref{fig:specgen}.}
    \label{fig:comsg3}
\end{figure}

An increasing IFD indicates the introduction of generalists in the team. When communication between specialists and generalists is not possible due to the high similarity threshold, agents still work individually, without communicating with each other. The increase of performance is due to generalists being able to solve all components of the task, and by increasing the ratio of generalists in the team, progressively more tasks are being solved by them. When teams are composed entirely of generalists (IFD = 1) all tasks are solved, and the time required for this drops dramatically (Figure \ref{fig:specgen} \textit{C}). Notably, communication is absent even in teams composed entirely of generalists, which is explained by the fact, that each generalist alone is capable of solving any task, and hence does not require communication. It is also worth noting, that in this system DFD does not have an effect on performance, time of task processing, or communication.

\begin{figure}[!h]
    \centering
    \includegraphics[width=\textwidth]{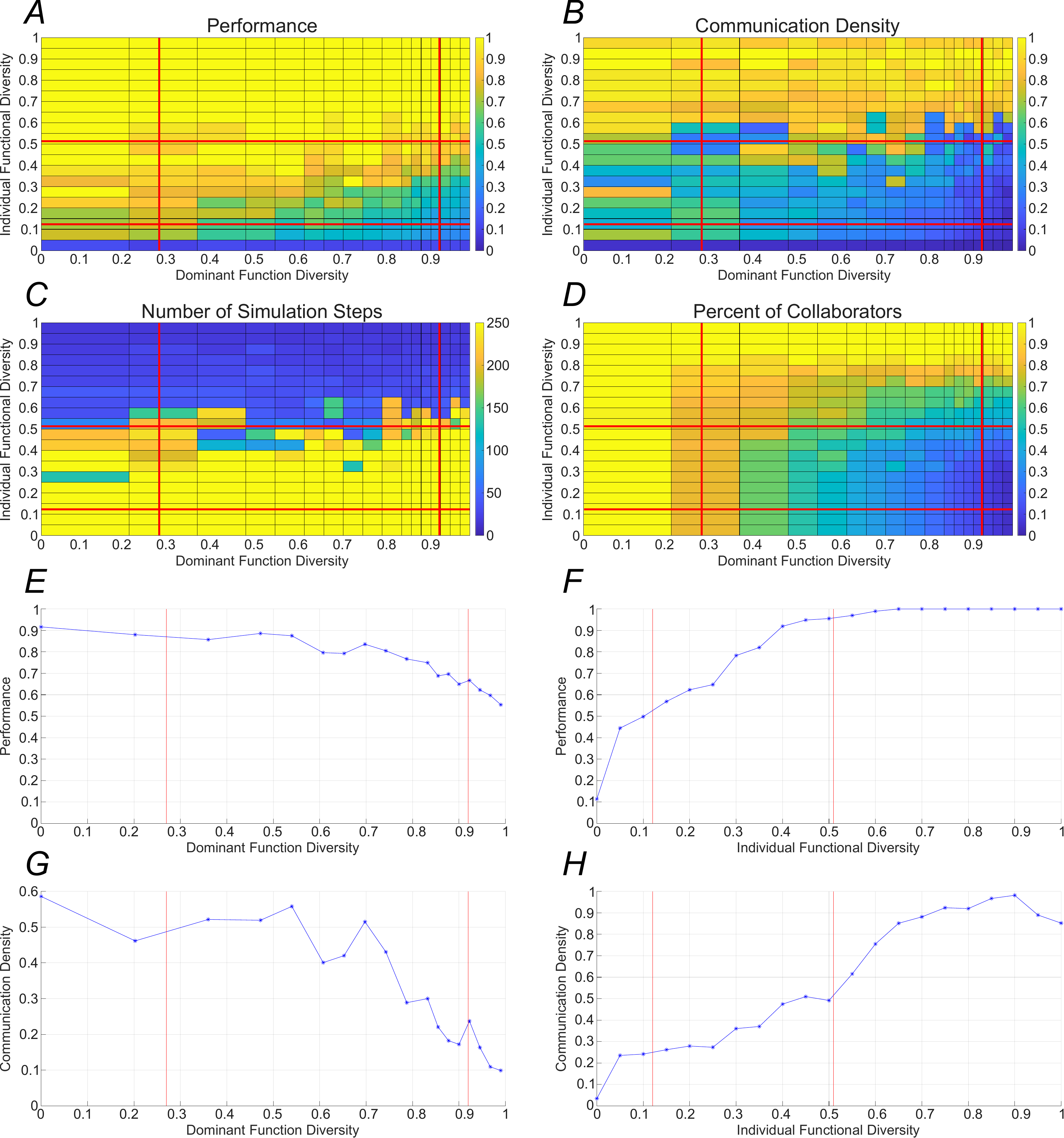}
    \caption{Simulation results of a team of diverse agents. Skills of agents are distributed according to a Gaussian distribution. Agents use the second communication scheme, i.e.\ they check if a more proficient collaborator is available in every time step. Figure panels \textit{A}–-\textit{D} are set up similarly to panels in Figure \ref{fig:specgen}. \textit{E}, \textit{F}: Performance averaged over IFD, and DFD, respectively. \textit{G}, \textit{H}: Communication density averaged over IFD, and DFD, respectively. Red vertical lines indicate the range of DFD, IFD values, respectively, available to empirical study by \citet{bunderson_comparing_2002}.}
    \label{fig:lackcomm}
\end{figure}

When a decreased similarity threshold allows for the formation of collaboration between specialists and generalists the team starts to perform well. Figure \ref{fig:comspecgen} shows team characteristics for the first communication scheme in which the passing of tasks occurs when an agent -- in the current simulation, a specialist -- gets stuck and passes their task to a generalist. For most IFD -- DFD values all tasks are completely solved (Figure \ref{fig:comspecgen} \textit{A}). However, performance is still poor when only specialists make up a team (i.e.\ IFD =0), since specialists of different functions are still too different to form collaborations, which blocks communication between specialists of different functions.

Communication density is a non-monotonous function of IFD (Figure \ref{fig:comspecgen} \textit{B}). In fact, communication density is highest when roughly half of the team is composed of specialists, and the other half is generalists. This is explained by a continuously decreasing number of task passing as the number of generalists increases (data not shown), and an also decreasing amount of time need to complete all tasks (Figure \ref{fig:comspecgen} \textit{C}). Interestingly enough, when agents repeatedly seek to find a more capable collaborator (i.e.\ using the second communication scheme), communication density plateaus due to a slowly decreasing number of task passing, and an initially rapidly dropping, then constant time needed to complete all tasks (Figure \ref{fig:comsg2} ).

When all agents are able to communicate with each other (Figure \ref{fig:comsg3}), most of the IFD -- DFD landscape of performance remain the same as in the above cases, however, since specialists of different function can also form collaborations, when the team is composed of specialists only, an increase in DFD results in increased communication and, in turn, increased performance.

\subsection{Does lack of communication degrade performance?}\label{ssec:comperf}
Most management teams, however, are not composed of pure specialists and absolute generalists. In a typical situation, each manager has some level of understanding of many functional areas. Consequently, their IFDS takes a real value between 0 and 1. Therefore, an ABM composed of agents with a distribution of skill strength values was used to simulate a team of more realistic problem solvers (Figure \ref{fig:TaskAgTeam} \textit{Cb}).

\begin{table}[h]
	\centering
	\begin{tabularx}{\textwidth}{lXXXl}
	\toprule
    \textbf{Communication density} & Estimate & SE & \textit{t}-statistics & \textit{p}-value \\
	\midrule
    Intercept & 0.28 & 0.03 & 10.25 & <0.001\\
    IFD & 0.96 & 0.03 & 34.49 & <0.001\\
    DFD & -0.28 & 0.03 & -9.01 & <0.001\\
    \cmidrule{2-5}
    & $df =375$ & adjusted $R^2 = 0.77$ & $F =635$ & $p< 0.001$\\
    \midrule
    \textbf{Performance} & Estimate & SE & \textit{t}-statistics & \textit{p}-value \\
    \hline
    Intercept & 40.59 & 1.60 & 25.37 & <0.001\\
    IFD & 49.79 & 1.66 & 29.97 & <0.001\\
    DFD & -12.62 & 1.85 & -6.81 & <0.001\\
    \cmidrule{2-5}
    & $df =375$ &  adjusted $R^2 =0.77$ &  $F =472$ & $p< 0.001$\\
    \bottomrule
	\end{tabularx}
	\caption{Dependence of communication density and performance on IFD and DFD. Multivariate linear regression modeling results for a group of diverse agents repeatedly seeking the best collaborator.}
	\label{tab:fit1}	
\end{table}

Similar to the previous case of pure specialists and absolute generalists, IFD has significant
influence on dependent measures. Using the second communication scheme, both the team’s performance (Figure \ref{fig:lackcomm} \textit{A}), and the amount of communication density (Figure \ref{fig:lackcomm} \textit{B}) increases significantly with increasing IFD (Table \ref{tab:fit1}), similar to previous findings \citep{bunderson_comparing_2002,zhou_functional_2023}.

\begin{figure}[h]
    \centering
    \includegraphics[width=\textwidth]{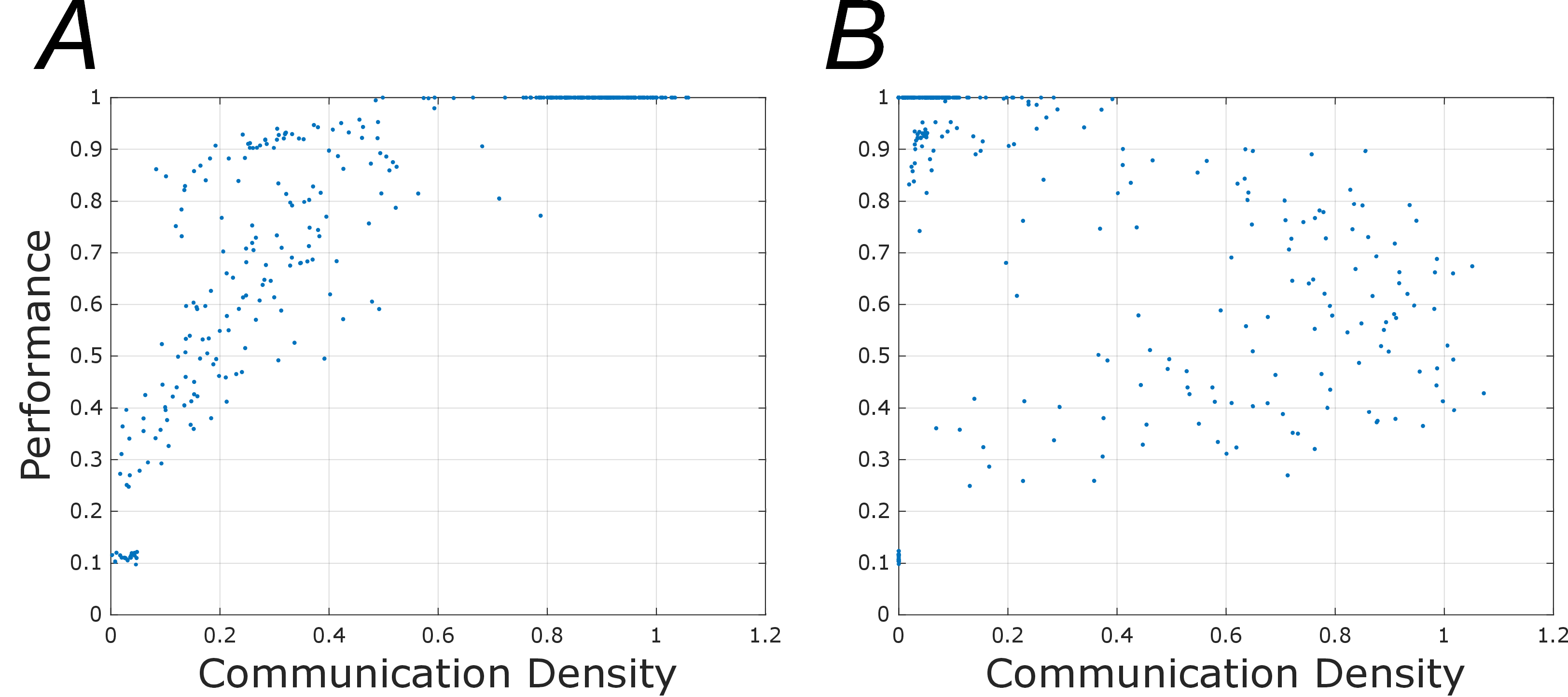}
    \caption{Performance as a function of communication density. When agents repeatedly seek to find the best collaborator (second communication scheme, \textit{A}), performance is significantly and strongly positively correlated with communication density ($r(376)=0.83$, $p<0.001$). However, when agents only communicate when they cannot proceed with solving their task (first communication scheme, \textit{B}), communication density significantly negatively correlates with performance ($r(376)=-0.48$, $p<0.001$).}
    \label{fig:percommcorr}
\end{figure}

A major difference compared to previous simulation results, however, is that additionally to their IFD dependence, performance and communication density became DFD-dependent as well in a large range of IFD values. Specifically, both measures decrease significantly when DFD increases as empirically determined by \citet{bunderson_comparing_2002} (Table \ref{tab:fit1}). Figure \ref{fig:lackcomm} \textit{E} \& \textit{G} highlight this trend by averaging performance and communication density for possible IFD values, and representing their dependence on DFD. Direct comparison of performance and communication density suggests that higher performance is related to more intensive communication, which is supported by correlation analysis showing strong and significant correlation (Figure \ref{fig:percommcorr} \textit{A}).

Changing communication style, and allowing agents to evaluate passing options only when they cannot proceed with solving their task, however, changes the communication density landscape (Figure \ref{fig:lackcomm2}). In particular, since it requires more time to complete tasks in this case relative to the previous case, communication density drops for high IFD values, resulting in a moderate and significantly negative correlation between performance and communication density
(Table \ref{tab:fit2}; Figure \ref{fig:percommcorr} \textit{B}).

\begin{table}[h]
	\centering
	\begin{tabularx}{\textwidth}{lXXXl}
	\toprule
    \textbf{Communication density} & Estimate & SE & \textit{t}-statistics & \textit{p}-value \\
	\midrule
    Intercept & 0.74 & 0.04 & 19.22 & < 0.001\\
    IFD &-0.74 & 0.04 & -18.33 & < 0.001\\
    DFD & -0.16 & 0.04 & -3.56 & < 0.001\\
    \cmidrule{2-5}
    & $df = 375$ & adjusted $R^2 = 0.48$ & $F = 175$ & $p < 0.001$\\
    \midrule
    \textbf{Performance} & Estimate & SE & \textit{t}-statistics & \textit{p}-value \\
    \hline
    Intercept & 38.71 & 1.58 & 24.44 & < 0.001\\
    IFD & 51.78 & 1.64 & 31.49 & < 0.001\\
    DFD & -12.24 & 1.83 & -6.67 & < 0.001\\
    \cmidrule{2-5}
    & $df = 375$ & adjusted $R^2 = 0.73$ & $F = 518$ & $p < 0.001$\\
    \bottomrule
	\end{tabularx}
	\caption{Multivariate linear regression modeling results of agents seeking collaborator help when they cannot proceed with solving their task.}
	\label{tab:fit2}
\end{table}

\subsection{Missing expertise is reflected in performance surface}
Referring to contingency theory, \citet{zhou_functional_2023} suggest that heterogeneous teams are better suited to solve unconventional and new problems than homogeneous ones. Hence, they reason that the management team of a newly founded small or medium enterprise (SME) with high level of DFD can more comprehensively understand and resolve complex issues in the current economic environment of China. They argue that compared to listed companies, new SMEs show higher uncertainty in terms of resources and rules. However, newly founded and just forming firms might face issues of missing expertise \citep{boeker_new_2005} hindering firm growth.

\begin{figure}[h]
    \centering
    \includegraphics[width=\textwidth]{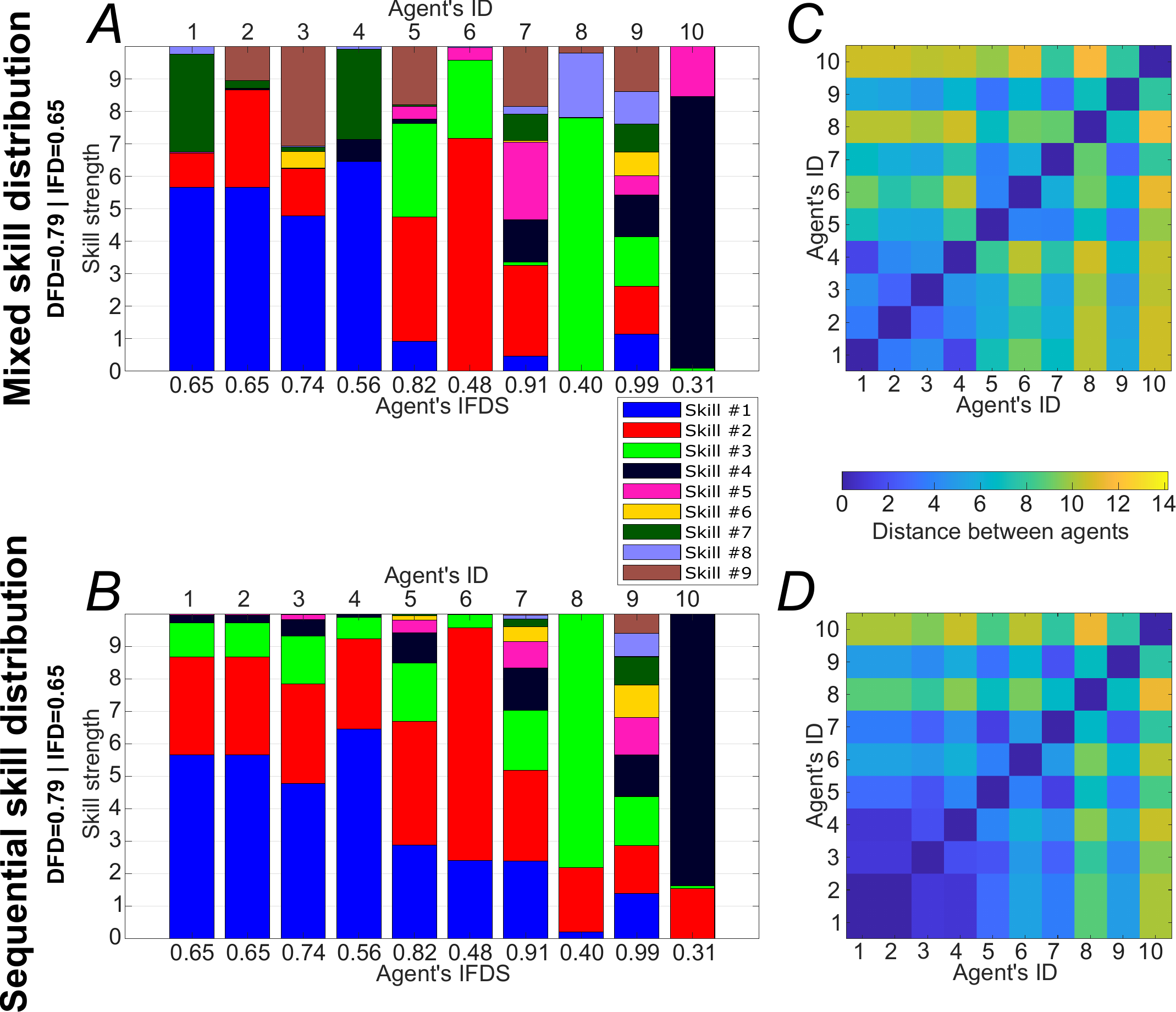}
    \caption{Teams of agents exemplifying missing expertise. Two teams of agents with identical diversity measures ($IFD = 0.65$, $DFD = 0.79$), but different coverage of necessary function set were generated to study performance and communication pattern differences. \textit{A}: after skill strength values are generated, the type of skill is randomly assigned to agents. \textit{B}: in this group of agents skill type is assigned to agents in a strictly increasing order. Note that in this group skills \#5 and up are very weakly represented compared to the team shown in panel \textit{A}. \textit{C}, \textit{D}: Distance between agents determining communication for the team shown in \textit{A} and \textit{B}, respectively.}
    \label{fig:missskillgroup}
\end{figure}

Well-established functional diversity measures, like IFD and DFD are expected to discriminate fully competent management teams from those that lack certain key competencies, but this is not always the case. To simulate the effect of missing expertise, i.e.\ incomplete skill coverage, teams of agents were created in the proposed ABM using two different methods. In the first method (Figure \ref{fig:missskillgroup} \textit{A}), used in previous examples, during generation of agents, dominant functions were assigned to them such that the DFD value specified for the simulation for a given team was attained. Remaining skill values were assigned to agents randomly, independent of the skill’s index. In contrast, the skill strengths of all agents in the second method (Figure \ref{fig:missskillgroup} \textit{B}) decreased with increasing skill index, subject to the defined DFD value. These agents than possess a higher degree of similarity compared to those generated by the first method. Specifically, in this group, for low DFD every agent could perform skill \#1 best, and their capabilities continuously decreased such that most of them were not competent in performing skill \#9. With increasing DFD, the dominant skill of some agents was changed from skill \#1 to skill \#2, but skill strength distribution was not changed otherwise, it was still decreasing with increasing skill index. Using these two methods, teams of agents with identical IFD--DFD values could be created, yet their collaboration networks (Figure \ref{fig:missskillgroup} \textit{C} \& \textit{D}), and more importantly, their coverage of the whole knowledge base was fundamentally different.

Simulations of teams of agents with missing expertise were conducted. Interestingly enough, Figure \ref{fig:msgperf} \textit{A} \& \textit{E} show that performance of these teams at low DFD values is low, but increases at high DFD. Compared to previous results (Figure \ref{fig:lackcomm} \textit{A} \& \textit{E}), performance in teams of agents missing necessary expertise is low for low DFD values but surpasses the performance of more heterogeneous agents at high DFD. These simulation results resemble empirical findings by \citet{zhou_functional_2023} in terms of exhibiting increasing performance both as a function of IFD, as well as of DFD (Table \ref{tab:fit3}).

\begin{table}[h]
	\centering
	\begin{tabularx}{\textwidth}{lXXXl}
	\toprule
    \textbf{Communication density} & Estimate & SE & \textit{t}-statistics & \textit{p}-value \\
	\midrule
    Intercept & 0.17 & 0.02 & 8.44 & <0.001\\
    IFD & 0.84 & 0.02 & 40.30 & <0.001\\
    DFD & -0.01 & 0.02 & -0.58 & 0.56\\
    \cmidrule{2-5}
    & $df = 375$ & adjusted $R^2 = 0.81$ & $F =812$ & $p< 0.001$\\
    \midrule
    \textbf{Performance} & Estimate & SE & \textit{t}-statistics & \textit{p}-value \\
    \hline
    Intercept & 18.62 & 1.58 & 11.75 & <0.001\\
    IFD & 63.66 & 1.64 & 38.71 & <0.001\\
    DFD & 3.36 & 1.84 & 1.83 & 0.06\\
    \cmidrule{2-5}
    & $df = 375$ & adjusted $R^2 = 0.77$ & $F =472 $ & $p< 0.001$\\
    \bottomrule
	\end{tabularx}
	\caption{Multivariate linear regression modeling results of a team of agents missing some necessary expertise.}
	\label{tab:fit3}
\end{table}

\subsection{Suggestion for a new measure of team diversity}
\begin{figure}[h]
    \centering
    \includegraphics[width=\textwidth]{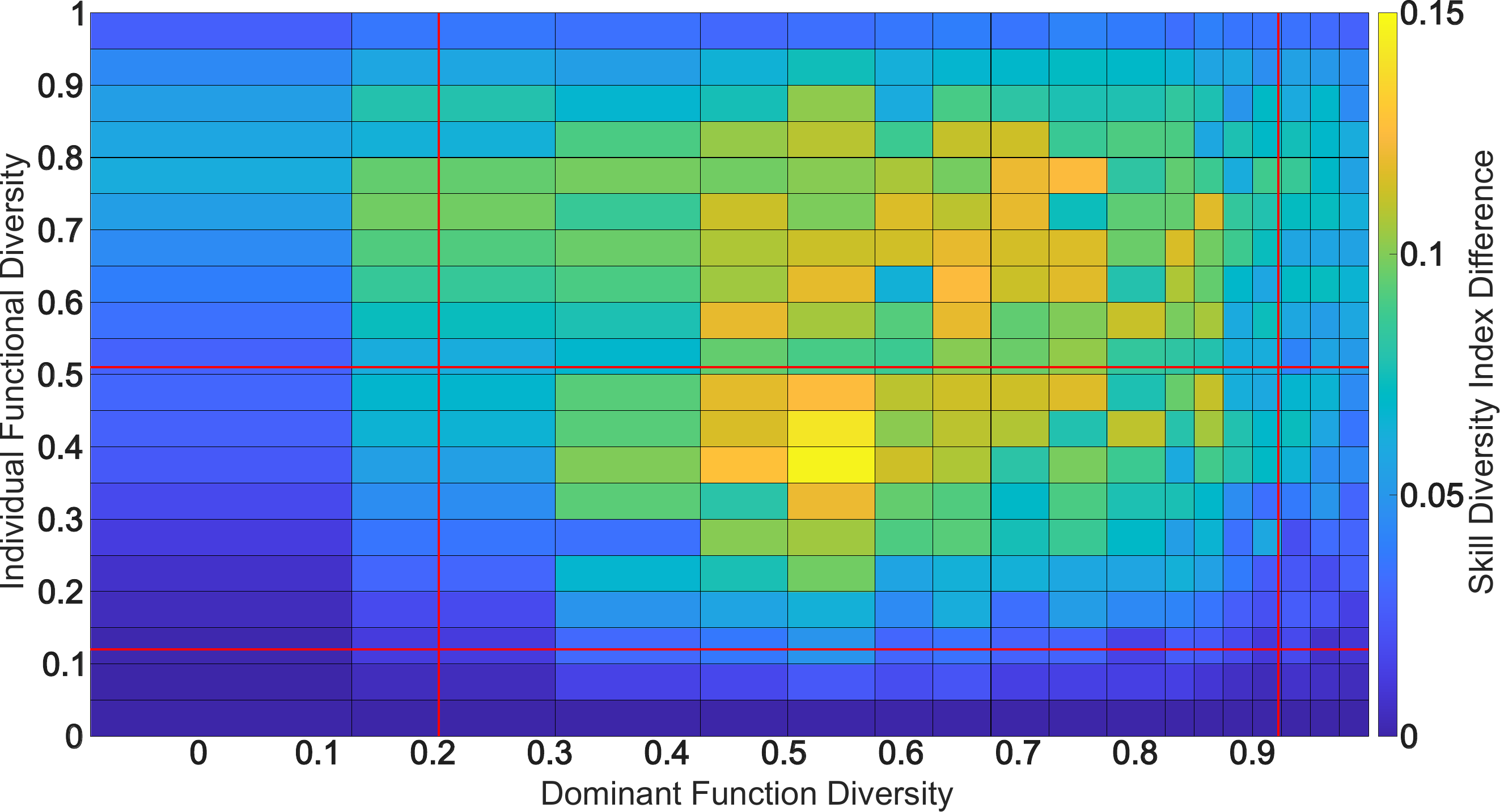}
    \caption{Difference of skill coverage for teams of equal IFD and DFD. Skill diversity indices were calculated for teams composed of agents with randomly distributed, as well as sequentially distributed skill types, and their difference plotted as a function of IFD and DFD.}
    \label{fig:SDI}
\end{figure}

The previous illustration shows that two teams of agents with identical IFD–-DFD values might perform differently because their skills do not cover necessary functional requirements equally. Thus, I recommend the introduction of another measure, called Skill Diversity Index or SDI, to account for a context dependent description of functional diversity:
\begin{equation}
    \mathit{SDI} = \frac{\left(1-\sum\limits_{j=1}^{N_\mathrm{Functions}} s_{j}^2 \right)}{1-\frac{1}{N_\mathrm{Functions}}}
    \label{eq:SDI}
\end{equation}
where $s_j$ indicates the amount of expertise all team members together have in performing function $j$, and $N_\mathrm{Functions}$ is the number of functions considered. Calculated analogously to IFD and DFD, SDI is a standardized metric with values spanning from 0 to 1. It assesses the distribution of skills across the entire team, with lower SDI values signifying a restricted range of function coverage, while SDI near 1 indicates that all functions are equally well-addressed. For example, Figure \ref{fig:SDI} shows an example of SDI difference for two sets of teams created by the two methods described in the previous section.

\begin{figure}[!h]
    \centering
    \includegraphics[width=\textwidth]{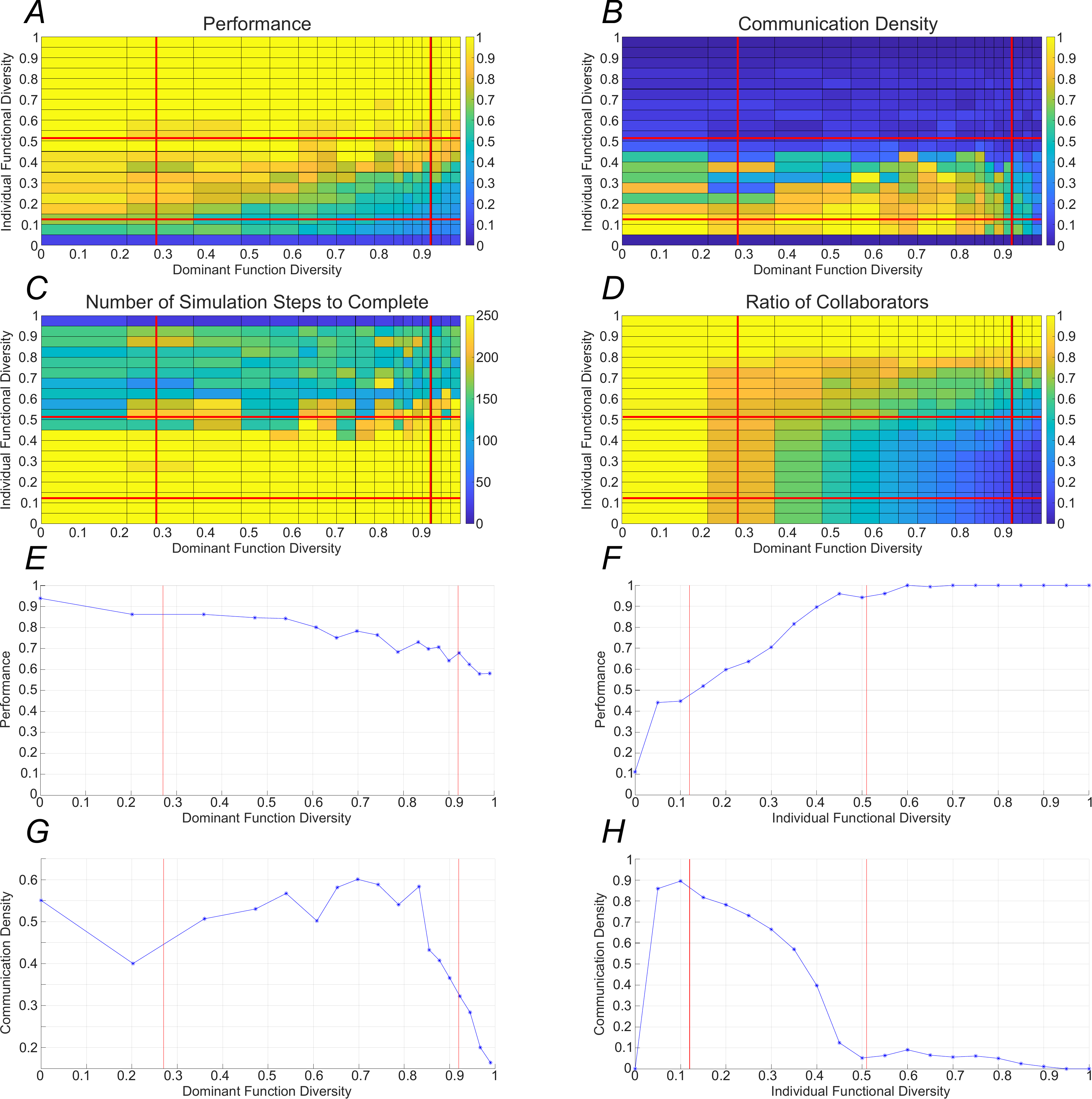}
    \caption{Simulation results of a team of agents seeking the best solver only when they cannot proceed with solving the task assigned to them. Simulation results might be compared with an analogous system represented in Figure \ref{fig:lackcomm}.}
    \label{fig:lackcomm2}
\end{figure}

\section{Discussion}\label{sec:discu}

This study introduces an agent-based model comprising agents, representing problem solvers or management team members, and tasks, representing projects composed of various functions. By analyzing increasingly complex scenarios using this model, we uncover a potential explanation for the contradictory empirical findings of \citet{bunderson_comparing_2002} and \citet{zhou_functional_2023} regarding DFD and management team performance. Additionally, we propose a novel metric, the Skill Diversity Index, to measure the extent to which team members collectively possess the skill sets required for effective management.

Specifically, we showed that generalists, i.e.\ team members with a broad range of functional experience, or a team composed of such generalists are more capable of solving a set of tasks than functional specialists. Based on analysis of the presented ABM, the reason for this is two-fold. On one hand, the broad functional experience of generalists allows them to find an aspect of the task at hand they are able to work on, and can more often proceed with solving it than specialists who are often blocked by not being able to find a task matching their skills. On the other hand, generalists are more similar to each other than specialists of different functions, and thus communicate easier with each other than dissimilar specialists. Indeed, while the possibility of a “double-edged sword” \citep{ma_top_2021,milliken_searching_1996} exists, most literature agree that a higher level of IFD corresponds to improved information sharing, and allows teams to be less susceptible to decision-making biases \citep{cannella_top_2008,smith_power_2006}. However, considering a more thorough exploration of the role of knowledge depth and knowledge width \citep{mannucci_differential_2018}, one might argue that higher IFD means that team members have spent less time performing individual functions, which makes it less likely that they have a deep understanding of the functions they have performed, resulting in superficial and unprofessional decision-making. In the model we accounted for this effect by normalizing skill vectors, which results in generalists being slower in solving specific functions than specialists. However -- in the current simulations at least -- communication advantages, and the broad knowledge base possessed by generalist outweigh their slowness in task processing.

Computational modeling allows for testing hypothesis regarding a specific component of a large system while keeping other components constant, a difficult scenario in empirical hypothesis testing. For example, we implemented two procedures by which agents might exchange information: in the first method agents communicate only if they need help because they cannot proceed with solving their task, while in the second, they constantly check for a more capable collaborator and if they find one, pass their task to them for further processing. Comparing these two ways of communication without modifying other aspects of the model yields interesting results: depending on the communication scheme, increased amount of communication might not always increase performance. Using the second communication scheme, when agents repeatedly evaluate the possibility of passing their task, increasing IFD increases communication (Figure \ref{fig:lackcomm} \textit{B}) by introducing agents possessing wider range of functional diversity. Since these agents are better in solving problems, performance increases, giving rise to a positive correlation between communication and performance (Figure \ref{fig:percommcorr} \textit{A}). Contrary to this, in a team of agents using the first communication scheme -- agents only communicate when they cannot further proceed with solving their task -- the amount of communication decreases with increasing IFD (Figure \ref{fig:lackcomm2}), since agents with a broader range of functional diversity most likely can proceed with their task and hence do not require communication. Thus, in this case there is a negative correlation between communication and performance (Figure \ref{fig:percommcorr} \textit{B}). Indeed, while information sharing is generally found to positively predict team performance \citep{mesmer-magnus_information_2009}, some studies suggest that positive impact of information sharing is not always the case \citep{xiao_does_2016}.

\begin{figure}[h]
    \centering
    \includegraphics[width=\textwidth]{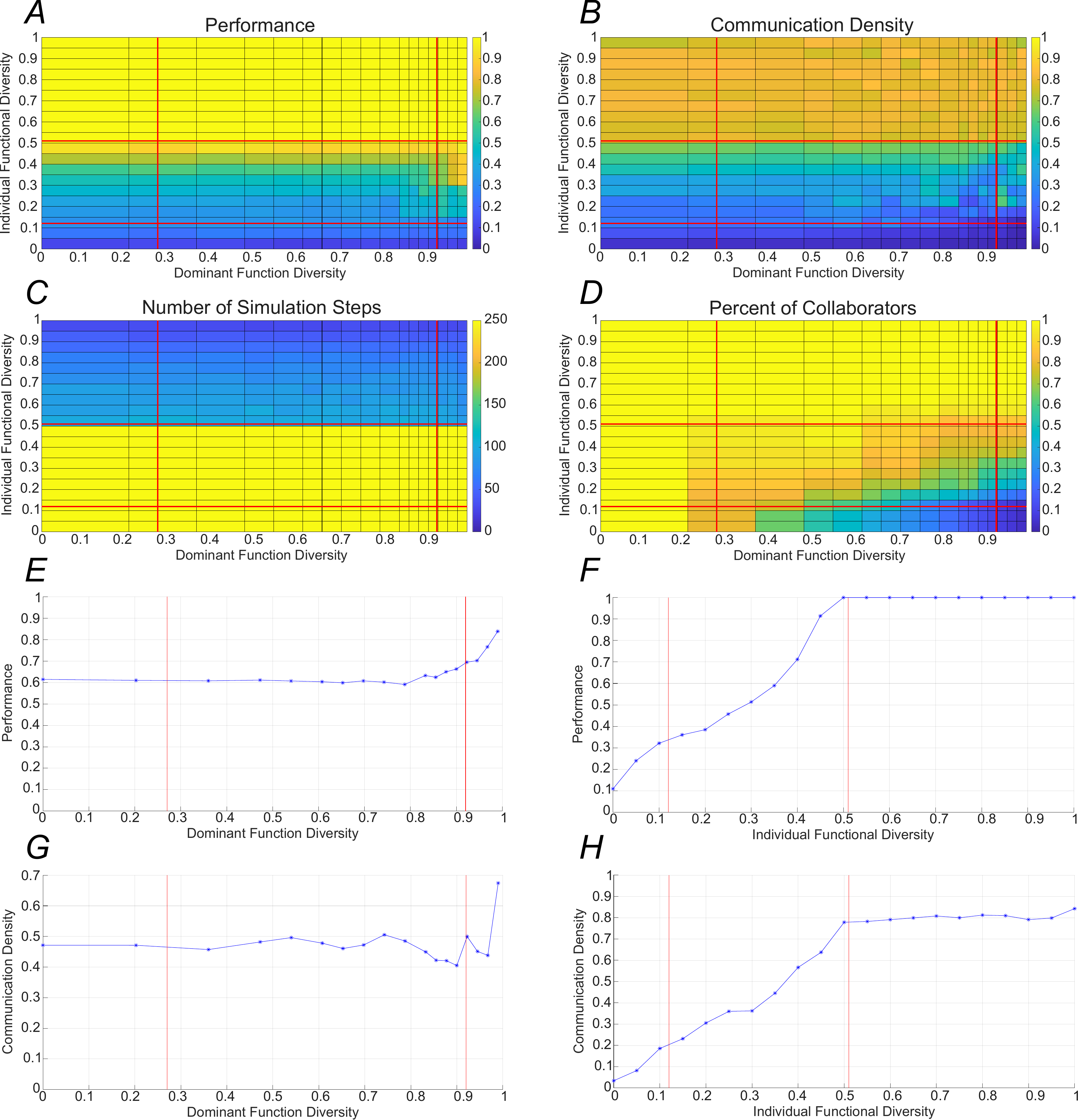}
    \caption{Simulation results for a team of agents that do not fully cover the range of required expertise. Performance in these teams increase with DFD (c.f.\ Figure \ref{fig:lackcomm}).}
    \label{fig:msgperf}
\end{figure}

Another goal of this study was to elaborate on the role DFD plays in influencing team performance. In terms of the effect of DFD on team performance, the literature is less unanimous than in the case of IFD. Evaluation of different effects of DFD has a long history with empirical research studying its connection with innovation \citep{bantel_top_1989} or consensus, and conflict \citep{knight_top_1999}. Most importantly, studies show that the effect of DFD depends on multiple factors and could influence near-term and long-term performance differently \citep{hambrick_influence_1996,murray_top_1989}. In particular, homogeneous management teams interact more effectively, but heterogeneous teams better facilitate adaptation and mitigate ``group thinking'' \citep{cannella_top_2008}. Conversely, high DFD means strong path dependence, which might increase conflict and makes communication difficult \citep{oreilly_iii_work_1989,stewart_meta-analytic_2006,williams_demography_1998}.

Performance might have different types of dependence on DFD due to differences in moderator variables, such as location, company size, date and method of data collection, etc. Also, the way performance is measured might largely influence experimental outcome. Indeed, there are considerable differences between the studies by \citet{bunderson_comparing_2002}, and \citet{zhou_functional_2023} in terms of both the moderator variables and the performance measurement methodology. For example, while \citet{bunderson_comparing_2002} studied 44 business unit management teams in a Fortune 100 consumer products company in the United States, with top management teams (TMTs) of about 11 members in the early 2000s, \citet{zhou_functional_2023} examined 762 small- and medium-sized Chinese enterprises 20 years later, that were founded less than eight years prior to their research, having TMTs of two to five members.  Also, \citeauthor{bunderson_comparing_2002} measured performance via a single variable, the profitability of the business unit, while \citeauthor{zhou_functional_2023} followed a multidimensional approach and recorded ten items to calculate firm performance.

Nevertheless, computational modeling can be used to fix many of the confounding factors and analyze the effect of a single variable independently. In the presented comparison all conditions were fixed, except for the coverage of necessary functions by the team’s overall skill set. Computer simulations showed that this difference resulted in an opposite DFD dependence of performance. In fact, a comparison of Figures \ref{fig:lackcomm} \textit{A} and \ref{fig:msgperf} \textit{A} shows that for moderate IFD, and low DFD values, the performance is lower when skills are missing (Figure \ref{fig:msgperf} \textit{A}) compared to the case when all necessary functions are covered by the team’s skill set (Figure \ref{fig:lackcomm} \textit{A}). As DFD increases, however, the tendency changes and teams with missing skills perform better, due to increased communication capability (c.f.\ Figures \ref{fig:lackcomm} \textit{B} and \ref{fig:msgperf} \textit{B}).

As previously mentioned, while projection type operations are valuable for reducing data dimensionality, their use carries inherent risks. This applies specifically to the maximum function used in defining DFD. \citet{bunderson_comparing_2002} outline DFD's link to a team's ability to handle essential functions: ``The extent to which the dominant functions of a team’s members are evenly distributed across all the relevant functions is viewed as an indication of the team’s breadth and balance of knowledge and expertise related to running all aspects of an organization'' However, our computer simulations revealed that DFD values alone do not determine skill coverage. They solely indicate which function a team member dominates without considering their depth of knowledge in that area. To bridge this gap in characterizing performance-oriented problem-solving teams, we proposed a new metric -- the Skill Diversity Index -- to evaluate how comprehensively a team collectively covers the necessary functions for operating an organization in its entirety.

Contemporary technology facilitates the swift gathering and analysis of extensive empirical data. The array of computational tools and data-sharing platforms available allows researchers to circumvent unnecessary simplification of multivariate data and present it in a more comprehensive manner than ever before. This study aimed to emphasize the significance of nuanced aspects, such as the distribution of skills among members of management teams, which, despite being collected, frequently vanish during the data analysis process. Additionally, it underscored the value of leveraging agent-based modeling, and defined a new measure of functional diversity with its help.

\bibliographystyle{unsrtnat}

\bibliography{ATRANwdoi} 

@article{nkomo_diversity_2019,
	title = {Diversity at a {Critical} {Juncture}: {New} {Theories} for a {Complex} {Phenomenon}},
	volume = {44},
	issn = {0363-7425, 1930-3807},
	shorttitle = {Diversity at a {Critical} {Juncture}},
	url = {http://journals.aom.org/doi/10.5465/amr.2019.0103},
	doi = {10.5465/amr.2019.0103},
	language = {en},
	number = {3},
	urldate = {2022-10-24},
	journal = {Academy of Management Review},
	author = {Nkomo, Stella M. and Bell, Myrtle P. and Roberts, Laura Morgan and Joshi, Aparna and Thatcher, Sherry M. B.},
	month = jul,
	year = {2019},
	pages = {498--517},
}

@inproceedings{yen_cast_2001,
	title = {{CAST}: {Collaborative} agents for simulating teamwork},
    booktitle={International Joint Conference on Artificial Intelligence},
	volume = {17},
	isbn = {1045-0823},
	publisher = {Lawrence Erlbaum Associates LTD},
	author = {Yen, John and Yin, Jianwen and Ioerger, Thomas R and Miller, Michael S and Xu, Dianxiang and Volz, Richard A},
	year = {2001},
	note = {Issue: 1},
	pages = {1135--1144}
}

@article{tambe_towards_1997,
	title = {Towards {Flexible} {Teamwork}},
	volume = {7},
	issn = {1076-9757},
	url = {https://jair.org/index.php/jair/article/view/10193},
	doi = {10.1613/jair.433},
	urldate = {2022-10-21},
	journal = {Journal of Artificial Intelligence Research},
	author = {Tambe, M.},
	month = sep,
	year = {1997},
	pages = {83--124},
}

@article{jennings_controlling_1995,
	title = {Controlling cooperative problem solving in industrial multi-agent systems using joint intentions},
	volume = {75},
	issn = {00043702},
	url = {https://linkinghub.elsevier.com/retrieve/pii/0004370294000202},
	doi = {10.1016/0004-3702(94)00020-2},
	language = {en},
	number = {2},
	urldate = {2022-10-21},
	journal = {Artificial Intelligence},
	author = {Jennings, N.R.},
	month = jun,
	year = {1995},
	pages = {195--240},
}

@article{tsvetovat_modeling_2004,
	title = {Modeling {Complex} {Socio}-technical {Systems} using {Multi}-{Agent} {Simulation} {Methods}},
	language = {en},
	number = {2},
    volume={18},
	journal = {Künstliche Intelligenz},
	author = {Tsvetovat, Maksim and Carley, Kathleen M},
	year = {2004},
	pages = {23--28},
}

@phdthesis{rojas-villafane_agent-based_2010,
	type = {Doctoral {D}issertation},
	title = {An {Agent}-based {Model} of {Team} {Coordination} and {Performance}},
	url = {https://digitalcommons.fiu.edu/etd/250},
	language = {en},
	urldate = {2022-10-21},
	school = {Florida International University},
	author = {Rojas-Villafane, Jose A},
	month = may,
	year = {2010},
	doi = {10.25148/etd.FI10081217},
}

@article{jin_virtual_1996,
	title = {The virtual design team: {A} computational model of project organizations},
	volume = {2},
	issn = {1381-298X, 1572-9346},
	shorttitle = {The virtual design team},
	url = {http://link.springer.com/10.1007/BF00127273},
	doi = {10.1007/BF00127273},
	language = {en},
	number = {3},
	urldate = {2022-10-21},
	journal = {Computational and Mathematical Organization Theory},
	author = {Jin, Yan and Levitt, Raymond E.},
	year = {1996},
	pages = {171--195},
}

@article{fan_modeling_2004,
	title = {Modeling and simulating human teamwork behaviors using intelligent agents},
	volume = {1},
	issn = {1571-0645},
	url = {http://www.scopus.com/inward/record.url?scp=10844262719&partnerID=8YFLogxK},
	doi = {10.1016/j.plrev.2004.10.001},
	number = {3},
	journal = {Physics of Life Reviews},
	author = {Fan, Xiaocong and Yen, John},
	month = dec,
	year = {2004},
	pages = {173--201},
}

@article{dehkordi_impacts_2012,
  title={Impacts of project-overload on innovation inside organizations: agent-based modeling},
  author={Dehkordi, Farnaz Motamediyan and Thompson, Anthony and Larsson, Tobias},
  journal={International Journal of Social, Behavioral, Educational, Economic, Business and Industrial Engineering},
  volume={6},
  number={11},
  pages={2808--2813},
  year={2012}
}

@inproceedings{perisic_agent-based_2016,
  address = {Dubrovnik, Hrvatska},
  title={Agent-based simulation framework to support management of teams performing development activities},
  author={Peri{\v{s}}i{\'c}, Marija Majda and Martinec, Tomislav and {\v{S}}torga, Mario and Kandu{\v{c}}, Tadej and others},
  booktitle={DS 84: Proceedings of the DESIGN 2016 14th International Design Conference},
  volume = {Sociotechnical Issues In Design},
  series = {Design},
  pages={1925--1936},
  issn={18479073},
  year={2016}
}

@article{fernandes_modelling_2017,
	title = {Modelling the dynamics of complex early design processes: an agent-based approach},
	volume = {3},
	issn = {2053-4701},
	shorttitle = {Modelling the dynamics of complex early design processes},
	url = {https://www.cambridge.org/core/product/identifier/S2053470117000178/type/journal_article},
	doi = {10.1017/dsj.2017.17},
	language = {en},
	urldate = {2022-10-19},
	journal = {Design Science},
	author = {Fernandes, João Ventura and Henriques, Elsa and Silva, Arlindo and Pimentel, César},
	year = {2017},
	pages = {e19},
}

@inproceedings{vermillion_using_2015,
	address = {Boston, Massachusetts, USA},
	title = {Using a {Principal}-{Agent} {Model} to {Investigate} {Delegation} in {Systems} {Engineering}},
	isbn = {978-0-7918-5705-2},
	url = {https://asmedigitalcollection.asme.org/IDETC-CIE/proceedings/IDETC-CIE2015/57052/Boston,%20Massachusetts,%20USA/256959},
	doi = {10.1115/DETC2015-47778},
	urldate = {2022-10-19},
	booktitle = {Volume {1B}: 35th {Computers} and {Information} in {Engineering} {Conference}},
	publisher = {American Society of Mechanical Engineers},
	author = {Vermillion, Sean D. and Malak, Richard J.},
	month = aug,
	year = {2015},
	pages = {V01BT02A046},
}

@book{taylor_agent-based_2014,
	series = {{OR} {Essentials}},
	title = {Agent-based {Modeling} and {Simulation}},
	isbn = {978-1-137-45364-8},
	url = {https://books.google.hu/books?id=UFlvBAAAQBAJ},
	publisher = {Palgrave Macmillan UK},
	author = {Taylor, S.},
	year = {2014},
}

@article{de_carolis_why_2009,
	title = {Why {Networks} {Enhance} the {Progress} of {New} {Venture} {Creation}: {The} {Influence} of {Social} {Capital} and {Cognition}},
	volume = {33},
	issn = {1042-2587},
	url = {https://doi.org/10.1111/j.1540-6520.2009.00302.x},
	doi = {10.1111/j.1540-6520.2009.00302.x},
	number = {2},
	urldate = {2022-10-10},
	journal = {Entrepreneurship Theory and Practice},
	author = {De Carolis, Donna Marie and Litzky, Barrie E. and Eddleston, Kimberly A.},
	month = mar,
	year = {2009},
	note = {Publisher: SAGE Publications Inc},
	pages = {527--545},
}

@article{jehn_why_1999,
	title = {Why {Differences} {Make} a {Difference}: {A} {Field} {Study} of {Diversity}, {Conflict} and {Performance} in {Workgroups}},
	volume = {44},
	issn = {0001-8392},
	url = {https://journals.sagepub.com/doi/abs/10.2307/2667054},
	doi = {10.2307/2667054},
	number = {4},
	urldate = {2022-10-07},
	journal = {Administrative Science Quarterly},
	author = {Jehn, Karen A. and Northcraft, Gregory B. and Neale, Margaret A.},
	month = dec,
	year = {1999},
	note = {Publisher: SAGE Publications Inc},
	pages = {741--763},
}

@article{cannella_top_2008,
	title = {Top {Management} {Team} {Functional} {Background} {Diversity} and {Firm} {Performance}: {Examining} the {Roles} of {Team} {Member} {Colocation} and {Environmental} {Uncertainty}},
	volume = {51},
	issn = {00014273},
	url = {http://www.jstor.org/stable/20159538},
	number = {4},
	urldate = {2022-09-29},
	journal = {The Academy of Management Journal},
	author = {Cannella, Albert A. and Park, Jong-Hun and Lee, Ho-Uk},
	year = {2008},
	note = {Publisher: Academy of Management},
	pages = {768--784},
}

@article{milliken_searching_1996,
	title = {Searching for {Common} {Threads}: {Understanding} the {Multiple} {Effects} of {Diversity} in {Organizational} {Groups}},
	volume = {21},
	issn = {03637425},
	url = {http://www.jstor.org/stable/258667},
	doi = {10.2307/258667},
	number = {2},
	urldate = {2022-09-29},
	journal = {The Academy of Management Review},
	author = {Milliken, Frances J. and Martins, Luis L.},
	year = {1996},
	note = {Publisher: Academy of Management},
	pages = {402--433},
}

@article{bunderson_comparing_2002,
	title = {Comparing alternative conceptualizations of functional diversity in management teams: {Process} and performance effects},
	volume = {45},
	language = {en},
	number = {5},
	journal = {Academy of Management Journal},
	author = {Bunderson, J Stuart and Sutcliffe, Kathleen M},
	year = {2002},
	pages = {875--893},
}

@article{williams_demography_1998,
	title = {Demography and {Diversity} in {Organizations}: {A} {Review} of 40 {Years} of {Research}},
	volume = {20},
	journal = {Research in organizational behavior},
	author = {Williams, Katherine Y and O'Reilly III, Charles A},
	month = jan,
	year = {1998},
	pages = {77--140},
}

@article{smith_top_1994,
	title = {Top {Management} {Team} {Demography} and {Process}: {The} {Role} of {Social} {Integration} and {Communication}},
	volume = {39},
	issn = {00018392},
	shorttitle = {Top {Management} {Team} {Demography} and {Process}},
	url = {https://www.jstor.org/stable/2393297?origin=crossref},
	doi = {10.2307/2393297},
	language = {en},
	number = {3},
	urldate = {2022-10-10},
	journal = {Administrative Science Quarterly},
	author = {Smith, Ken G. and Smith, Ken A. and Olian, Judy D. and Sims, Henry P. and O'Bannon, Douglas P. and Scully, Judith A.},
	month = sep,
	year = {1994},
	pages = {412},
}

@article{aboramadan_top_2021,
	title = {Top management teams characteristics and firms performance: literature review and avenues for future research},
	volume = {29},
	issn = {1934-8835, 1934-8835},
	shorttitle = {Top management teams characteristics and firms performance},
	url = {https://www.emerald.com/insight/content/doi/10.1108/IJOA-02-2020-2046/full/html},
	doi = {10.1108/IJOA-02-2020-2046},
	language = {en},
	number = {3},
	urldate = {2022-10-07},
	journal = {International Journal of Organizational Analysis},
	author = {Aboramadan, Mohammed},
	month = may,
	year = {2021},
	pages = {603--628},
}

@article{van_knippenberg_work_2004,
	title = {Work {Group} {Diversity} and {Group} {Performance}: {An} {Integrative} {Model} and {Research} {Agenda}.},
	volume = {89},
	issn = {1939-1854, 0021-9010},
	shorttitle = {Work {Group} {Diversity} and {Group} {Performance}},
	url = {http://doi.apa.org/getdoi.cfm?doi=10.1037/0021-9010.89.6.1008},
	doi = {10.1037/0021-9010.89.6.1008},
	language = {en},
	number = {6},
	urldate = {2022-10-07},
	journal = {Journal of Applied Psychology},
	author = {van Knippenberg, Daan and De Dreu, Carsten K. W. and Homan, Astrid C.},
	month = dec,
	year = {2004},
	pages = {1008--1022},
}

@article{hambrick_upper_1984,
	title = {Upper {Echelons}: {The} {Organization} as a {Reflection} of {Its} {Top} {Managers}},
	volume = {9},
	issn = {0363-7425},
	url = {https://doi.org/10.5465/amr.1984.4277628},
	doi = {10.5465/amr.1984.4277628},
	number = {2},
	urldate = {2022-10-07},
	journal = {Academy of Management Review},
	author = {Hambrick, Donald C. and Mason, Phyllis A.},
	month = apr,
	year = {1984},
	note = {Publisher: Academy of Management},
	pages = {193--206},
}

@article{homberg_top_2013,
	title = {Top {Management} {Team} {Diversity}: {A} {Systematic} {Review}},
	volume = {38},
	issn = {1059-6011, 1552-3993},
	shorttitle = {Top {Management} {Team} {Diversity}},
	url = {http://journals.sagepub.com/doi/10.1177/1059601113493925},
	doi = {10.1177/1059601113493925},
	language = {en},
	number = {4},
	urldate = {2022-10-07},
	journal = {Group \& Organization Management},
	author = {Homberg, Fabian and Bui, Hong T. M.},
	month = aug,
	year = {2013},
	pages = {455--479},
}

@incollection{jackson_consequences_1996,
	address = {UK},
	edition = {1st},
	title = {The {Consequences} of {Diversity} in {Multidisciplinary} {Work} {Teams}},
	isbn = {0471957909},
	language = {en},
	booktitle = {Handbook of work group psychology},
	publisher = {John Wiley \& Sons Ltd},
	author = {Jackson, Susan E},
	year = {1996},
	pages = {25},
}

@article{zhou_functional_2023,
	title = {Functional diversity of top management teams and firm performance in {SMEs}: a social network perspective},
	issn = {1863-6683, 1863-6691},
	shorttitle = {Functional diversity of top management teams and firm performance in {SMEs}},
	url = {https://link.springer.com/10.1007/s11846-022-00524-w},
	doi = {10.1007/s11846-022-00524-w},
	language = {en},
    volume = {17},
    pages ={259--286},
	urldate = {2022-09-28},
	journal = {Review of Managerial Science},
	author = {Zhou, Lulu and Huang, Haiyan and Chen, Xiaolin and Tian, Feng},
	month = jan,
	year = {2023},
}

@article{hong_groups_2004,
	title = {Groups of diverse problem solvers can outperform groups of high-ability problem solvers},
	volume = {101},
	issn = {0027-8424, 1091-6490},
	url = {https://pnas.org/doi/full/10.1073/pnas.0403723101},
	doi = {10.1073/pnas.0403723101},
	language = {en},
	number = {46},
	urldate = {2022-03-29},
	journal = {Proceedings of the National Academy of Sciences},
	author = {Hong, Lu and Page, Scott E.},
	month = nov,
	year = {2004},
	pages = {16385--16389},
}

@inproceedings{chang_differential_2012,
	address = {Malacca, Malaysia},
	title = {Differential effects of knowledge diversity on team innovation: {An} agent-based modeling},
	isbn = {978-1-4673-0654-6 978-1-4673-0655-3 978-1-4673-0653-9},
	shorttitle = {Differential effects of knowledge diversity on team innovation},
	url = {http://ieeexplore.ieee.org/document/6236384/},
	doi = {10.1109/ICIMTR.2012.6236384},
	language = {en},
	urldate = {2022-03-21},
	booktitle = {2012 {International} {Conference} on {Innovation} {Management} and {Technology} {Research}},
	publisher = {IEEE},
	author = {Chang, Wan-Jing},
	month = may,
	year = {2012},
	pages = {179--182},
}

@article{bonabeau_agent-based_2002,
	title = {Agent-based modeling: {Methods} and techniques for simulating human systems},
	volume = {99},
	issn = {0027-8424, 1091-6490},
	shorttitle = {Agent-based modeling},
	url = {https://pnas.org/doi/full/10.1073/pnas.082080899},
	doi = {10.1073/pnas.082080899},
	language = {en},
	number = {suppl\_3},
	urldate = {2022-03-21},
	journal = {Proceedings of the National Academy of Sciences},
	author = {Bonabeau, Eric},
	month = may,
	year = {2002},
	pages = {7280--7287},
}

@article{walsh_selectivity_1988,
	title = {SELECTIVITY AND SELECTIVE PERCEPTION: {An} INVESTIGATION OF MANAGERS' BELIEF STRUCTURE AND INFORMATION PROCESSING.},
	volume = {31},
	issn = {0001-4273, 1948-0989},
	shorttitle = {{SELECTIVITY} {AND} {SELECTIVE} {PERCEPTION}},
	url = {http://amj.aom.org/cgi/doi/10.2307/256343},
	doi = {10.2307/256343},
	language = {en},
	number = {4},
	urldate = {2023-05-08},
	journal = {Academy of Management Journal},
	author = {Walsh, J. P.},
	month = dec,
	year = {1988},
	pages = {873--896},
}

@article{boeker_new_2005,
	title = {New {Venture} {Evolution} and {Managerial} {Capabilities}},
	volume = {16},
	issn = {1047-7039, 1526-5455},
	url = {https://pubsonline.informs.org/doi/10.1287/orsc.1050.0115},
	doi = {10.1287/orsc.1050.0115},
	language = {en},
	number = {2},
	urldate = {2023-05-12},
	journal = {Organization Science},
	author = {Boeker, Warren and Wiltbank, Robert},
	month = apr,
	year = {2005},
	pages = {123--133},
}

@article{smith_power_2006,
	title = {Power relationships among top managers: {Does} top management team power distribution matter for organizational performance?},
	volume = {59},
	issn = {01482963},
	shorttitle = {Power relationships among top managers},
	url = {https://linkinghub.elsevier.com/retrieve/pii/S0148296305001670},
	doi = {10.1016/j.jbusres.2005.10.012},
	language = {en},
	number = {5},
	urldate = {2023-05-19},
	journal = {Journal of Business Research},
	author = {Smith, Anne and Houghton, Susan M. and Hood, Jacqueline N. and Ryman, Joel A.},
	month = may,
	year = {2006},
	pages = {622--629},
}

@article{ma_top_2021,
	title = {Top {Management} {Team} {Intrapersonal} {Functional} {Diversity} and {Adaptive} {Firm} {Performance}: {The} {Moderating} {Roles} of the {CEO}–{TMT} {Power} {Gap} and {Severity} of {Threat}},
	volume = {12},
	issn = {1664-1078},
	shorttitle = {Top {Management} {Team} {Intrapersonal} {Functional} {Diversity} and {Adaptive} {Firm} {Performance}},
	url = {https://www.frontiersin.org/articles/10.3389/fpsyg.2021.772739/full},
	doi = {10.3389/fpsyg.2021.772739},
	language = {en},
	urldate = {2023-05-19},
	journal = {Frontiers in Psychology},
	author = {Ma, Changlong and Ge, Yuhui and Wang, Jingwei},
	month = dec,
	year = {2021},
	pages = {772739},
}

@article{mannucci_differential_2018,
	title = {The {Differential} {Impact} of {Knowledge} {Depth} and {Knowledge} {Breadth} on {Creativity} over {Individual} {Careers}},
	volume = {61},
	issn = {0001-4273, 1948-0989},
	url = {http://journals.aom.org/doi/10.5465/amj.2016.0529},
	doi = {10.5465/amj.2016.0529},
	language = {en},
	number = {5},
	urldate = {2023-05-19},
	journal = {Academy of Management Journal},
	author = {Mannucci, Pier Vittorio and Yong, Kevyn},
	month = oct,
	year = {2018},
	pages = {1741--1763},
}

@article{xiao_does_2016,
	title = {Does information sharing always improve team decision making? {An} examination of the hidden profile condition in new product development},
	volume = {69},
	issn = {01482963},
	shorttitle = {Does information sharing always improve team decision making?},
	url = {https://linkinghub.elsevier.com/retrieve/pii/S0148296315002258},
	doi = {10.1016/j.jbusres.2015.05.014},
	language = {en},
	number = {2},
	urldate = {2023-05-19},
	journal = {Journal of Business Research},
	author = {Xiao, Yazhen and Zhang, Haisu and Basadur, Timothy M.},
	month = feb,
	year = {2016},
	pages = {587--595},
}

@article{mesmer-magnus_information_2009,
	title = {Information sharing and team performance: {A} meta-analysis.},
	volume = {94},
	issn = {1939-1854, 0021-9010},
	shorttitle = {Information sharing and team performance},
	url = {http://doi.apa.org/getdoi.cfm?doi=10.1037/a0013773},
	doi = {10.1037/a0013773},
	language = {en},
	number = {2},
	urldate = {2023-05-19},
	journal = {Journal of Applied Psychology},
	author = {Mesmer-Magnus, Jessica R. and DeChurch, Leslie A.},
	year = {2009},
	pages = {535--546},
}

@article{bantel_top_1989,
	title = {Top management and innovations in banking: {Does} the composition of the top team make a difference?},
	volume = {10},
	issn = {01432095, 10970266},
	shorttitle = {Top management and innovations in banking},
	url = {https://onlinelibrary.wiley.com/doi/10.1002/smj.4250100709},
	doi = {10.1002/smj.4250100709},
	language = {en},
	number = {S1},
	urldate = {2023-05-19},
	journal = {Strategic Management Journal},
	author = {Bantel, Karen A. and Jackson, Susan E.},
	year = {1989},
	pages = {107--124},
}

@article{knight_top_1999,
	title = {Top management team diversity, group process, and strategic consensus},
	volume = {20},
	issn = {0143-2095, 1097-0266},
	url = {https://onlinelibrary.wiley.com/doi/10.1002/(SICI)1097-0266(199905)20:5<445::AID-SMJ27>3.0.CO;2-V},
	doi = {10.1002/(SICI)1097-0266(199905)20:5<445::AID-SMJ27>3.0.CO;2-V},
	language = {en},
	number = {5},
	urldate = {2023-05-19},
	journal = {Strategic Management Journal},
	author = {Knight, Don and Pearce, Craig L. and Smith, Ken G. and Olian, Judy D. and Sims, Henry P. and Smith, Ken A. and Flood, Patrick},
	month = may,
	year = {1999},
	pages = {445--465},
}

@article{murray_top_1989,
	title = {Top {Management} {Group} {Heterogeneity} and {Firm} {Performance}},
	volume = {10},
	url = {http://www.jstor.org/stable/2486586},
	language = {en},
	journal = {Strategic Management Journal},
	author = {Murray, Alan I.},
	year = {1989},
	pages = {125--141},
}

@article{hambrick_influence_1996,
	title = {The {Influence} of {Top} {Management} {Team} {Heterogeneity} on {Firms}' {Competitive} {Moves}},
	volume = {41},
	issn = {00018392},
	url = {https://www.jstor.org/stable/2393871?origin=crossref},
	doi = {10.2307/2393871},
	language = {en},
	number = {4},
	urldate = {2023-05-19},
	journal = {Administrative Science Quarterly},
	author = {Hambrick, Donald C. and Cho, Theresa Seung and Chen, Ming-Jer},
	month = dec,
	year = {1996},
	pages = {659},
}

@article{stewart_meta-analytic_2006,
	title = {A {Meta}-{Analytic} {Review} of {Relationships} {Between} {Team} {Design} {Features} and {Team} {Performance}},
	volume = {32},
	issn = {0149-2063, 1557-1211},
	url = {http://journals.sagepub.com/doi/10.1177/0149206305277792},
	doi = {10.1177/0149206305277792},
	language = {en},
	number = {1},
	urldate = {2023-05-23},
	journal = {Journal of Management},
	author = {Stewart, Greg L.},
	month = feb,
	year = {2006},
	pages = {29--55},
}

@article{oreilly_iii_work_1989,
	title = {Work {Group} {Demography}, {Social} {Integration}, and {Turnover}},
	volume = {34},
	issn = {00018392},
	url = {https://www.jstor.org/stable/2392984?origin=crossref},
	doi = {10.2307/2392984},
	language = {en},
	number = {1},
	urldate = {2023-05-23},
	journal = {Administrative Science Quarterly},
	author = {O'Reilly III, Charles A. and Caldwell, David F. and Barnett, William P.},
	month = mar,
	year = {1989},
	pages = {21},
}

@article{hirschman_paternity_1964,
	title = {The {Paternity} of an {Index}},
	volume = {54},
	url = {http://www.jstor.org/stable/1818582},
	language = {en},
	number = {5},
	journal = {The American Economic Review},
	author = {Hirschman, Albert O.},
	year = {1964},
	pages = {761},
}

\end{document}